\documentclass[twocolumn]{aastex62}
\usepackage{graphicx}	% Including figure files
\usepackage{amsmath}	% Advanced maths commands
\usepackage{amssymb}	% Extra maths symbols
\usepackage{physics}
\usepackage{enumitem}
\usepackage{bbm}
\usepackage{color}
\usepackage{natbib}
\newcommand{\kms}{$\rm{km \,s^{-1}}$}

\begin{document}
%\title{Active Galactic Nuclei are more common in real bulges than in pseudobulges}
\title{The Activation of Galactic Nuclei and Their Accretion Rates are Linked to the Star Formation Rates and Bulge-types of Their Host Galaxies}
\shorttitle{Random Forest bulge classification}
\shortauthors{Yesuf et al.}
\author{Hassen M. Yesuf}
\affiliation{UCO/Lick Observatory, Department of Astronomy and Astrophysics, University of California, Santa Cruz, CA 95064 USA}
\affiliation{Kavli Institute for Astronomy and Astrophysics, Peking University, Beijing 100871, China}
\affiliation{Kavli Institute for the Physics and Mathematics of the Universe, The University of Tokyo, Kashiwa, Japan 277-8583}
\author{S. M. Faber}
\affiliation{UCO/Lick Observatory, Department of Astronomy and Astrophysics, University of California, Santa Cruz, CA 95064 USA}
\author{David C. Koo}
\affiliation{UCO/Lick Observatory, Department of Astronomy and Astrophysics, University of California, Santa Cruz, CA 95064 USA}
\author{Joanna Woo}
\affiliation{Department of Physics \& Astronomy, PO Box 1700 STN CSC, Victoria BC V8W 2Y2, Canada}
\author{Joel R. Primack}
\affiliation{UCO/Lick Observatory, Department of Astronomy and Astrophysics, University of California, Santa Cruz, CA 95064 USA}
\affiliation{SCIPP, University of California, Santa Cruz, CA 95064 USA}
\author{Yifei Luo}
\affiliation{UCO/Lick Observatory, Department of Astronomy and Astrophysics, University of California, Santa Cruz, CA 95064 USA}

\begin{abstract}

We use bulge-type classifications of 809 representative SDSS galaxies by Gadotti (2009) to classify a large sample of galaxies into real bulges (classical or elliptical) and pseudobulges using Random Forest. We use structural and stellar population predictors that can easily be measured without image decomposition. Multiple parameters such as the central mass density with 1 kpc, concentration index, S\'{e}rsic index and velocity dispersion do result in accurate bulge classifications when combined together. We classify $\sim 44,500$ face-on galaxies above stellar mass of 10$^{10}$ M$_\odot$ and redshift $ 0.02 < z < 0.07$ into real bulges or pseudobulges with $93 \pm 2$\% accuracy. We show that  $\sim 75 - 90\%$ of AGNs identified by the optical line ratio diagnostic are hosted by real bulges. The pseudobulge fraction significantly decreases with AGN signature as the line ratios change from indicating pure star formation ($\sim 54 \pm 4$ \%), to composite of star formation and AGN ($\sim 18 \pm 3$\%), and to AGN-dominated galaxies ($\sim 5 \pm 3$\%). Using the dust-corrected [\ion{O}{3}] luminosity as an AGN accretion indicator, and the stellar mass and radius as proxies for a black hole mass, we find that AGNs in real bulges have lower Eddington ratios than AGNs in pseudobulges. Real bulges have a wide range of AGN and star formation activities, although most of them are weak AGNs. For both bulge-types, their Eddington ratios are correlated with specific star formation rates (SSFR). Real bulges have lower specific accretion rate but higher AGN fraction than pseudobulges do at similar SSFRs.\\

\end{abstract}


\section{INTRODUCTION} \label{sec:int}

The disk component of a galaxy is well described by an exponential surface brightness profile. A bulge is a high-density central structure that is brighter than the inward extrapolation of the exponential disk profile \citep[e.g.,][]{Freeman70, Fisher+16}, and it is not associated with a bar. Bulges can be identified using bulge-disk image decomposition techniques \citep[e.g.,][]{Allen+06,Gadotti09,Weinzirl+09,Simard+11}. These techniques may be prone to large uncertainties and give rise to differences in bulge-type classification among different authors. Some authors have put considerable efforts into producing more accurate results, for example, by adding bars in the image decomposition \citep{Gadotti09,Weinzirl+09}.

Bulges have been roughly classified into two categories: pseudobulges and classical bulges \citep[see review by][]{Kormendy+04,Fisher+16}. There is no single ideal way of defining these two categories. In general, compared to classical bulges, pseudobulges are more rotation-dominated, have less concentrated surface brightness profiles, and tend to show younger stellar populations and ongoing star formation. Pseudobulge comprises two distinct stellar structures: box-peanut (BP) bulges and disc-like bulges (or inner discs). BP bulges are the inner parts of bars that grow out of the disc plane, whereas inner discs are built from gas inflow to the center and  subsequent star  formation \citep{Athanassoula05}. In this study, we refer to both types as pseudobulges. The empirical evidence that there is more than one type of bulge may reflect different mechanisms of bulge formation and galaxy evolution. 

Currently, the mechanisms proposed for build-up of bulges include galaxy mergers \citep[e.g.,][]{Toomre+72,Hopkins+09,Hopkins+10}, slow secular evolutions such as bars \citep{Kormendy+04}, and violent clumpy disk instabilities \citep[e.g.,][]{Elmegreen+08, Dekel+09, Inoue+12,Bournaud+16}. Although our current theoretical understanding of bulges is incomplete \citep[for theoretical review of bulge formation see][]{Brooks+16}, the common view is that classical bulges are formed mainly by dissipative mergers of galaxies, while pseudobulges are formed by disk related secular processes internal to galaxies. Some pseudobulges may also originate from mergers or non-secular (rapid) dynamical processes \citep[e.g.,][]{Keselman+12,Guedes+13}.

Current models for bulge formation based on $\Lambda$CDM hierarchical build-up of structure are challenged even with existing empirical bulge classification based on small local samples \citep{Weinzirl+09,Peebles+10, Kormendy+10, Fisher+11}. In the models, major mergers are ubiquitous and readily produce classical bulges \citep{Somerville+15, Brooks+16}. But fewer classical bulges are observed than predicted by the models. \citet{Hopkins+09} found that including the effects of gas properly in mergers leads to better agreement, and as a result less galaxies are predicted to be bulge-dominated. \citet{Porter+14}, however, found that including a disk instability mechanism for spheroid formation, tuned to reproduce the abundances of spheroid-dominated galaxies, underpredicts the fraction of observed discs with small bulges \citep[$B/T < 0.2$,][]{Weinzirl+09}.

 \citet{Kormendy+10} studied nearby giant field galaxies with circular velocity $v_{\rm circ} > 150$ \kms\, within 8 Mpc of our Galaxy. They found that at least 11 out of 19 galaxies show no evidence for classical bulges. Almost all of the classical bulges that they identified are smaller than those of typical simulated galaxies. Only four of the giant galaxies are ellipticals or have classical bulges that contribute about a third of their total mass. If galaxy mergers are expected to happen frequently and pseudobulges are not primarily produced by mergers, \citet{Kormendy+10} argued that most of these giant galaxies would have classical bulges.

\citet{Fisher+11} studied bulge-types in a volume-limited sample within 11 Mpc using Spitzer 3.6 $\mu$m and Hubble Space Telescope (HST) data. They found that whether counting by number, mass, or star formation, the dominant galaxy types in the local universe are pseudobulges or bulgeless galaxies. They showed that these galaxies account for over 80\% of galaxies above a stellar mass of $10^{9}$\,M$_\odot$. The frequency of bulge-types strongly depends on galaxy mass. Bulgeless and pseudobulge galaxies are the most frequent types below stellar mass of $10^{10}$ M$_\odot$. Majority of their galaxies above $10^{10.5}$ M$_\odot$ are either ellipticals or classical bulges. 

Using $\sim 100-300$ galaxies with HST imaging, \citet{Fisher+08,Fisher+10,Fisher+16} showed that the bulge S\'{e}rsic indices, $n_b$, is bimodal, and this bimodality correlates with the morphology of the bulge. About 90\% of their pseudobulges have $n_b < 2$, and similar percentage of classical bulges have $n_b > 2$. Therefore, $n_b <  2$ is  a good, albeit imperfect, criterion to separate pseudobulges from classical bulges and elliptical galaxies when a high quality imaging is available. The $n_b$ measurements are uncertain by about 0.5  in low resolution images \citep{Gadotti09}. The physical basis for the $n_b =  2$ threshold is not well understood.

\citet{Gadotti09} studied structural properties of a larger sample of about 1000 galaxies from the Sloan Digital Sky Survey (SDSS). He argued that, while $n_b$ can be used to distinguish pseudobulges from classical bulges, a more reliable (for low resolution imaging), and physically motivated, separation can be made using the Kormendy relation \citep[see also][]{Neumann+17}. He defined pseudobulges as galaxies that lie 3$\sigma$ below this linear relation of the mean effective bulge surface brightness and the effective radius of the ellipticals. This definition was criticized by \citet{Fisher+16}, who showed that a significant population of their bright pseudobulges with $n_b < 2$ lie on the Kormendy relation. Since there is no single ideal way of identifying bulge-types, \citet{Fisher+16} and \citet{Neumann+17} called for a comprehensive approach that combines multiple indicators. Nevertheless, after carefully decomposing the SDSS multiband images into bulge, disc, and bar components, \citet{Gadotti09} showed that the Petrosian concentration index is a better proxy for the bulge-to-total ratio than the global S\'{e}rsic index. He also found that that 32\%  of the total stellar mass in massive galaxies ($ >10^{10}$\,M$_\odot$) at redshift $z < 0.07$ is contained in ellipticals, 36\% in discs, 25\% in classical bulges, and 3\% in pseudobulges.

By comparing with \citet{Gadotti09}'s bulge classification, \citet{Luo+19} study the relationship of the central mass density within 1 kpc ($\Sigma_1$) to the nature of bulges using a large sample from the SDSS. They find that the residual of $\Sigma_1$ after the mass trend is removed can be used to separate pseudobulges from classical bulges. In addition, they argue that  the non-linear relation between $\Sigma_1$ and the star formation rate may explain the discrepancies between bulge indicators in previous studies. They note the existence of a population of galaxies with high central densities which are classified structurally as classical bulges but have high star formation rates, like pseudobulges.

In this paper, we combine multiple bulge indicators, including $\Sigma_1$ and concentration index, using Random Forest machine learning algorithm. We accurately predict whether a galaxy has a pseudobulge or a real bulge using the \citet{Gadotti09}'s sample to train the algorithm. We use structural and stellar population predictors that can easily be measured without resorting to careful image decomposition employed in the training sample. We aim to demonstrate that machine learning is useful in classifying bulge-types of much larger samples of galaxies, and it can accurately recover existing definition of bulge-types. The machine classification provides better statistics for checking the consistency of $\Lambda$CDM hierarchical clustering theory of galaxy formation with observed pseudobulge frequencies based on a large sample of galaxies. In addition,  as a demonstration of the kind of new science that this new approach enables, we study the connection between AGN optical line ratios and the pseudobulge fraction, and whether AGNs in pseudobulges have different accretion luminosities than those of AGNs in real bulges.

The paper is structured as as follows: section~\ref{sec:data} describes the SDSS data, and the methods used in the paper. Section~\ref{sec:res} presents the results. Section~\ref{sec:disc} presents general discussions and implications. The main conclusions and the summary of the results are presented in Section~\ref{sec:conc}. A cosmology with $\Omega_m =0.27, \Omega_\Lambda =0.73$, and $H_0=70$ km s$^{-1}$ Mpc$^{-1}$ is assumed.

\section{DATA AND METHOD}\label{sec:data}

\subsection{Collating stellar and structural parameters}

We use the Sloan Digital Sky Survey \citep[SDSS][]{York+00,Alam+15}. The publicly available Catalog Archive Server (CAS)\footnote{http://skyserver.sdss.org/casjobs/} is used to retrieve and collate some of the measurements used in this work (e.g., stellar masses, star formation rates, WISE flux ratios, and spectral indices). These data are supplemented with structural parameters (S\'{e}rsic index, and ellipticity/axial ratio) derived from a single S\'{e}rsic fits, given in Table 3 of \citet{Simard+11}. 

We also use the central stellar mass surface density within 1 kpc, $\Sigma_1$, and  the half-mass radius measured by \citet{Woo+19}. Both of these quantities are computed from the extinction-corrected \textit{ugriz} surface brightness profiles. To compute the stellar mass profiles, the \textit{ugriz} SEDs within each profile bin were fitted with synthetic SEDs from the PEGASE 2 \citep{Fioc+99} stellar population synthesis code to obtain a best-fitting r-band mass to light ratio. $\Sigma_1$ is then computed by interpolating the cumulative stellar mass profile to 1 kpc.  Similarly, the half-mass radius is obtained by interpolating the cumulative mass profile to the radius that contains half the total mass. This total mass is also computed by SED fitting the object's total \textit{ugriz} (Petrosian) magnitudes (k-corrected using the code of  \citet{Blanton+07}).  For more details, see \citet{Woo+19}.

We define global effective mass density as $\Sigma = M_\star/(2\pi R_{1/2}^2)$, where $M_\star$ is the total stellar mass, and $R_{1/2}$ is  the half-mass radius computed from the mass profile. The concentration index is defined as the ratio of 90\% Petrosian r-band light radius to 50\% Petrosian radius, $C_r = R_{90}/R_{50}$. The 12 predictors we use to train the Random Forest algorithm are: $\Sigma_1$, $C_r$, global $\Sigma$, global S\'{e}rsic index, the 4000{\AA} break index ($D_n(4000))$, H$\delta$ absorption equivalent width, Lick Mgb index, the WISE 12\,$\mu$m flux to  4.6\,$\mu$m flux ratio ($\log f_{12}/f_{4.6}$), the fiber velocity dispersion ($\sigma$), the total stellar mass, the fiber stellar mass, and the total star formation rate (SFR). Due to the limited resolution of the SDSS images, the $\Sigma_1$ measurements are reliable only below $z < 0.075$ \citep{Fang+13}. Furthermore, because the SDSS fiber encloses larger areas of galaxies with redshift, the fiber stellar mass, velocity dispersion and the spectral indices may change significantly with redshift. For these reasons and for the fact that the training sample has $z < 0.07$, it is not advisable to apply our bulge classification technique to SDSS data beyond $z > 0.07$.

We exclude type 1 AGN from our sample. We include only type 2 AGNs by restricting the Balmer lines to have velocity dispersion less than 300 km s$^{-1}$ \citep{Oh+15}. For these objects the AGN has no significant effect on the measurements of the host galaxy properties. Table~\ref{tbl:samp_cut} summarizes the sample selection.

\begin{deluxetable*}{lccc}
\tabletypesize{\footnotesize}
\tablecolumns{4} 
\tablewidth{0pt} 
\tablecaption{Sample selection \label{tbl:samp_cut}}
\tablehead{
\colhead{Cut description} & \colhead{Criterion} & \colhead{Number of galaxies} & \colhead{Sample name}}
\startdata
		Redshift limit & $z = 0.02-0.07$& 156,229$^\dagger$  \\ 
	         Mass limit & $\log M$(M$_\odot) = 10-12$  & 80,960 & Main sample \\ 
	         Face-on (ellipticity cut)  & $e = 1-b/a < 0.5$ & 49,390\\ 
	         Not Broad line AGN & Balmer line width $\sigma < 300$ km s$^{-1}$ & 44,491 & \\
	         \hline
	         BPT line ratios & S/N $> 2$ for the four emission lines  & 32,839 & \\ 
	         AGN &  Above \citet{Kewley+01} curve & 10,949 & All AGNs \\ 
	         LINER & AGN and below \citet{Schawinski+07} line & 8,182 & LINER AGNs or LINERs\\ 
	         Seyfert  &  AGN and above \citet{Schawinski+07} line  & 2,747 & Seyfert AGNs or Seyferts\\ 
	         Composite &  Between \citet{Kauffmann+03}'s \& & 9,341 & AGN+SF composites or composites\\ 
	           & \citet{Kewley+01}'s curves &  & \\ 
	         star-forming &  Below \citet{Kauffmann+03}'s curve & 12,649 & Star-forming (SF)\\ 
	         \hline
	          & $z = 0.02-0.07$& & \\ 
	         Gadotti (2009)'s cuts & $\log M$ (M$_\odot) = 10-12$ & & \\ 
	          & $e = 1-b/a < 0.1$ & 809$^\dagger$ & Gadotti (2009)'s sample \\ 
	         %\hline
		  &  80\% of Gadotti (2009)'s sample & 647 &  Training sample \\ 
		  &  20\% of Gadotti (2009)'s sample &162 &  Test sample \\ 
		\enddata
	        \tablecomments{$\dagger$ After matching all the catalogs used in this paper (\S 2.1).}
\end{deluxetable*}

%We have not included the total stellar mass, the half-mass radius, the velocity dispersion and the fiber stellar mass as predictors because they correlate with black hole mass, and we aim to independently infer the AGN properties of the two bulge-types.

\subsection{Training and test sample}

We use \citet{Gadotti09}'s sample as training and test datasets. This volume-limited sample was selected from SDSS DR2, applying the following criteria: 1) redshift $0.02 < z < 0.07$. 2) stellar mass above $10^{10}$ M$_\odot$. 3) axial ratio $b/a \ge 0.9$. After accounting for the effects of the axial ratio cut, \citet{Gadotti09} argued that this sample is fairly representative of massive galaxies and AGNs in the local universe. We use 809 galaxies ($\sim 86$\%) in his sample, which have measurements of all 12 predictors used by Random Forest. We use 647 galaxies for training and validating the Random Forest algorithm and 162 galaxies for testing the algorithm. In our prediction of the bulge-types, we restrict the whole SDSS sample to the same redshift and mass ranges but we consider face-on galaxies with $b/a > 0.5$. 

\citet{Gadotti09} defined pseudobulges as galaxies that lie 3$\sigma$ below the Kormendy relation of the ellipticals. A complementary approach is to use bulge S\'{e}rsic indices $n_b < 2$ to define pseudobulges \citep{Fisher+16}. But measuring $n_b$ accurately may require high quality data. There is not a large sample of high resolution Fisher-Drory type sample for the SDSS. We choose to adopt \citet{Gadotti09}'s original classification.  We will discuss the sensitivity of our main results if we adopt the classification based on $n_b$ instead.

%This definition was criticized as less robust by \citet{Fisher+16}, who showed that a significant population of bright pseudobulges with bulge S\'{e}rsic indices $n_b < 2$ lie on the Kormendy relation. We choose to retain his original classification. Thus, our classifications will agree well with what Gadotti would get for the same galaxies, if he analyzed them. They might not agree with the classifications of Fisher and Drory. This kind of discrepancy is inevitable right now since there is not a large sample of high resolution Fisher-Drory standardized sample for the SDSS. Given the limitation of the current training sample, we will discuss the sensitivity of our main results if we adopt the classification based on $n_b$ instead. Less than 2\% of the galaxies in the training sample are classified as bulgeless galaxies. For the sake of simplicity, we do not attempt to make a distinction between bulgeless and pseudobulge galaxies,  and we rather group them together. We refer to his classical bulge and elliptical classes together as real bulges. We do not separate these two classes in our prediction because the training sample size is small.

\subsection{Random Forest}\label{sec:RF}

% Breiman, ?Random Forests?, Machine Learning, 45(1), 5-32, 2001.

Random Forest (RF) is a supervised machine learning algorithm \citep{Breiman01}. It grows an ensemble of classification (or regression) trees on bootstrapped training subsets of categorical (or continuous) datasets, and aggregates the classifications of the bootstrapped trees to improve the prediction accuracy and avoid over-fitting (i.e., learning complex patterns from simply generated but noisy data). When used for a classification, RF uses class votes or probabilistic predictions from each tree, and assigns a final class using a majority vote or a mean class probability of the trees in the forests. The RF algorithm further improves the variance of the predictions by randomly choosing maximum of $m$ out of $p$ predictors ($m = 5$ and $p = 12$ in our case) and picking the best variable among the $m$ predictors to split a node into two during a tree-growing process. The randomized selection of predictors at each split decorrelates the bootstrapped trees, thereby reducing the variance of a prediction and making it more reliable. %$m \sim \sqrt p$ is typically used as a default value for classification. 

\begin{figure*}
\includegraphics[width=0.98\linewidth]{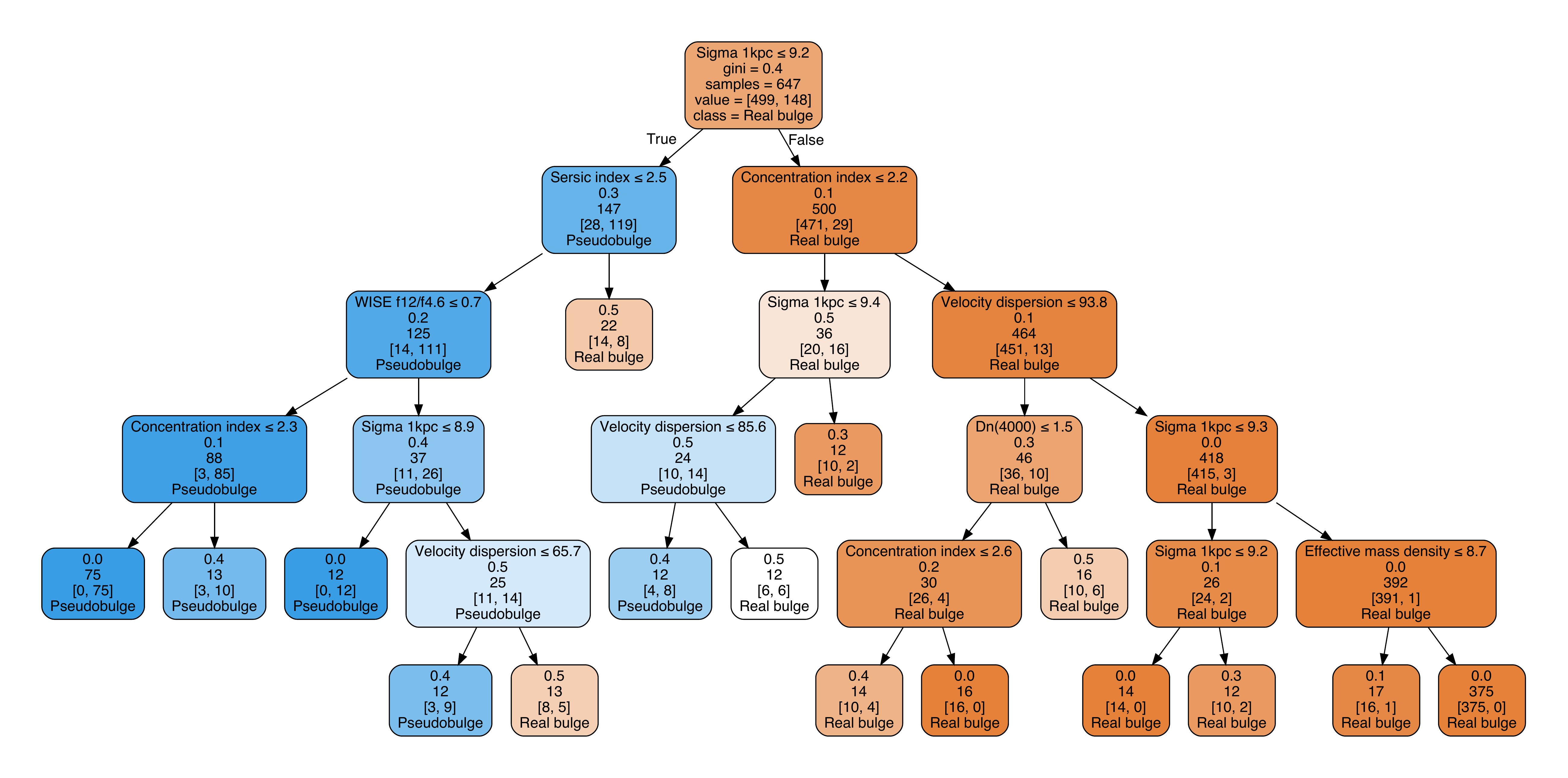}
\caption{An example decision tree for the training sample. A Random Forest is constructed by averaging a multitude of such decision trees. On a given node (box), the split criterion, the Gini index (a tree splitting metric), the sample size prior to a split, the number of galaxies in the two classes after the given split criteria, and the dominant class are shown. For example, in the root node (at the very top), the whole training sample (647) is divided in to real bulges (499) and pseudobulges (148) using a central mass density within 1 kpc, $\log\,\Sigma_1$, threshold of 9.2 M$_\odot$ kpc$^{-2}$ with the Gini index of 0.4. This index is zero when all galaxies satisfying a given criterion belong to the same class. Blue indicates a pseudobulge node and orange indicates a real bulge node. The darker the color, the less contaminated a given node is. \label{fig:tree}}
\end{figure*}

We use the Python \texttt{scikit-learn} library to implement RF \citep{scikit}. Using a ten fold cross-validation, we select the best hyper-parameters of the algorithm shown in bold from the sets: {\texttt{n\_estimators}: [50, 100, 200, \textbf{300}, 500], \texttt{max\_depth}: [\textbf{None}, 2, 3, 4, 5], \texttt{min\_samples\_leaf}: [6, 8, 10, \textbf{12}, 14, 16], and \texttt{max\_features}: [3, 4, \textbf{5}]}. This means that the best model grows 300 different trees. Each tree is grown by considering a maximum of 5 out of the 12 variables randomly to best split a node using the Gini impurity criterion until all leaves with more than twelve samples are pure or they reach the minimum sample of twelve. The \texttt{scikit-learn} implementation of RF combines classifiers by averaging their probabilistic predictions, and the final predicted class is the one with highest mean probability. For illustrative purpose, Figure~\ref{fig:tree} shows a single decision tree for the training sample.

%The 8 predictors used to construct the tree are : the mass density within 1 kpc, $\Sigma_1$, the concentration index, $C_r$, global $\Sigma$, global S\'{e}rsic index, the 4000{\AA} break index ($D_n(4000)$, H$\delta$ absorption index, Lick Mgb index, and the WISE 12\,$\mu$m flux to  4.6\,$\mu$m flux ratio ($\log f_{12}/f_{4.6}$).
%Explain variable importance,

\subsection{Error estimate of the pseudobulge fraction}

\citet{Rahardja+15} developed easy-to-implement maximum likelihood point and interval estimators for the true binomial proportion parameter using a double-sampling scheme. In this scheme, a small special sample of galaxies, for example, is classified by an infallible classifier (careful bulge-disk decomposition), but a fallible classifier (Random Forest) is only available to practically classify a large main sample of galaxies with some misclassification errors. The special and main sample need to be independent, and the observed number of objects classified by the fallible classifier is assumed to follow a binomial distribution. We use 20\% of the \citet{Gadotti09}'s sample as a test set to assess how well Random Forest recovers his classification, and to estimate the corresponding predicted mean pseudobulge fraction and the 95\% modified Wald confidence interval \citep{Rahardja+15}. We construct a main sample that does not contain the test sample, and applied Random Forest to both samples.

\subsection{Dust correction}

To use \ion{O}{3} 5007 luminosity as an AGN accretion indicator, with the dust attenuation correction based on the H$\alpha$/H$\beta$ ratio and the emission-line extinction curve given below \citep{Charlot+00,Wild+11,Yesuf+14}. If the observed H$\alpha$/H$\beta$ ratio of an AGN is less than 3.1 \citep[e.g.,][]{Ferland+83,Gaskell+84}, its \ion{O}{3} luminosity is not dust-corrected.

\begin{equation}
 Q_\lambda = 0.6\,(\lambda/5500)^{-1.3} + 0.4\,(\lambda/5500)^{-0.7}
\end{equation}

The optical depth at 5007{\AA} relative to the V band is :
\begin{equation}
 \tau_{5007} = \tau_V Q_{5007}
\end{equation}

\begin{equation}
 \tau_{V} = 0.921 \times \frac{2.5}{(Q_{4861} - Q_{6563})} \times \log \mathrm{\frac{H\alpha/H\beta}{3.1}}
\end{equation}

The dust corrected (dc) \ion{O}{3} luminosity is: 
\begin{equation}
L_{{\rm O3, dc}} = L_{{\rm O3}} \times 10^{0.4 \times 1.086 \tau_{5007}}
\end{equation}

\begin{figure*}
\centering
\includegraphics[width=0.90\linewidth]{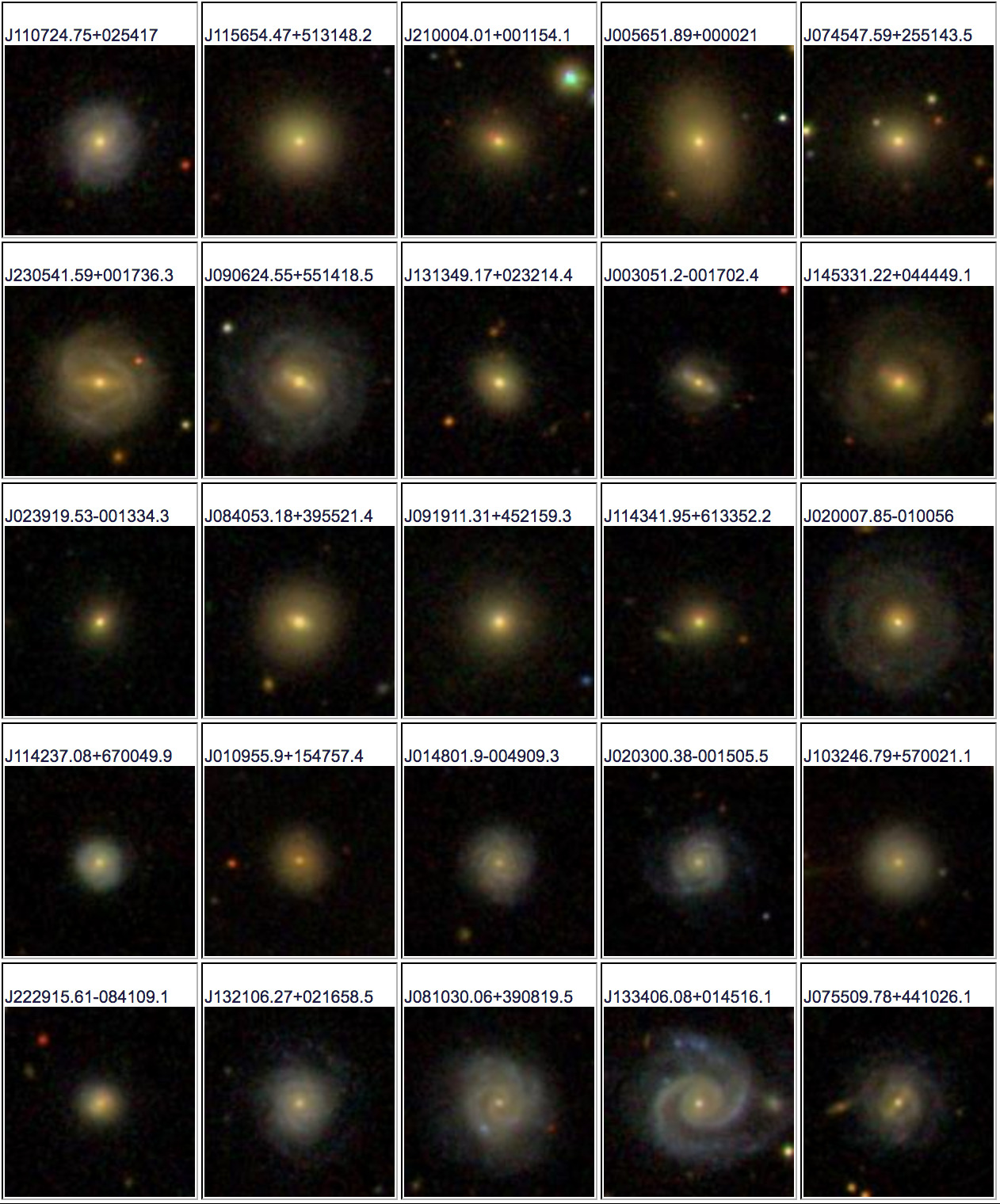}
\caption{Images of random galaxies from the training sample \citep{Gadotti09} correctly classified by the Random Forest algorithm. The top three rows are real bulges and the bottom two rows are pseudobulges. \label{fig:image_rf}}
\end{figure*}

\begin{figure*}
\includegraphics[width=0.90\textwidth]{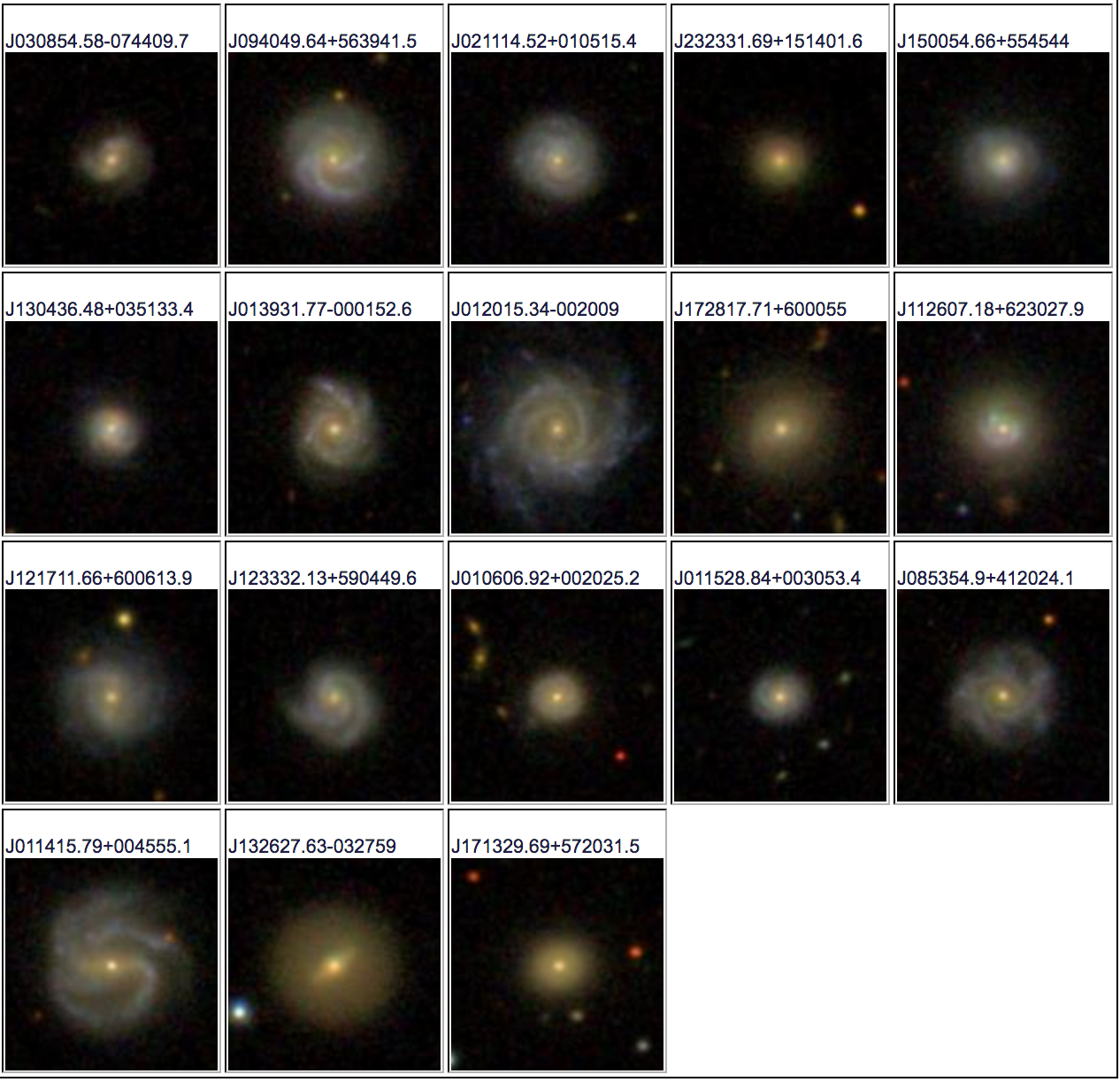}
\caption{Few galaxies from \citet{Gadotti09}'s sample that Random Forest classified as pseudobulges but \citet{Gadotti09} classified as real bulges. \label{fig:img_misRB}}
\end{figure*}

\begin{figure*}
\includegraphics[width=0.90\textwidth]{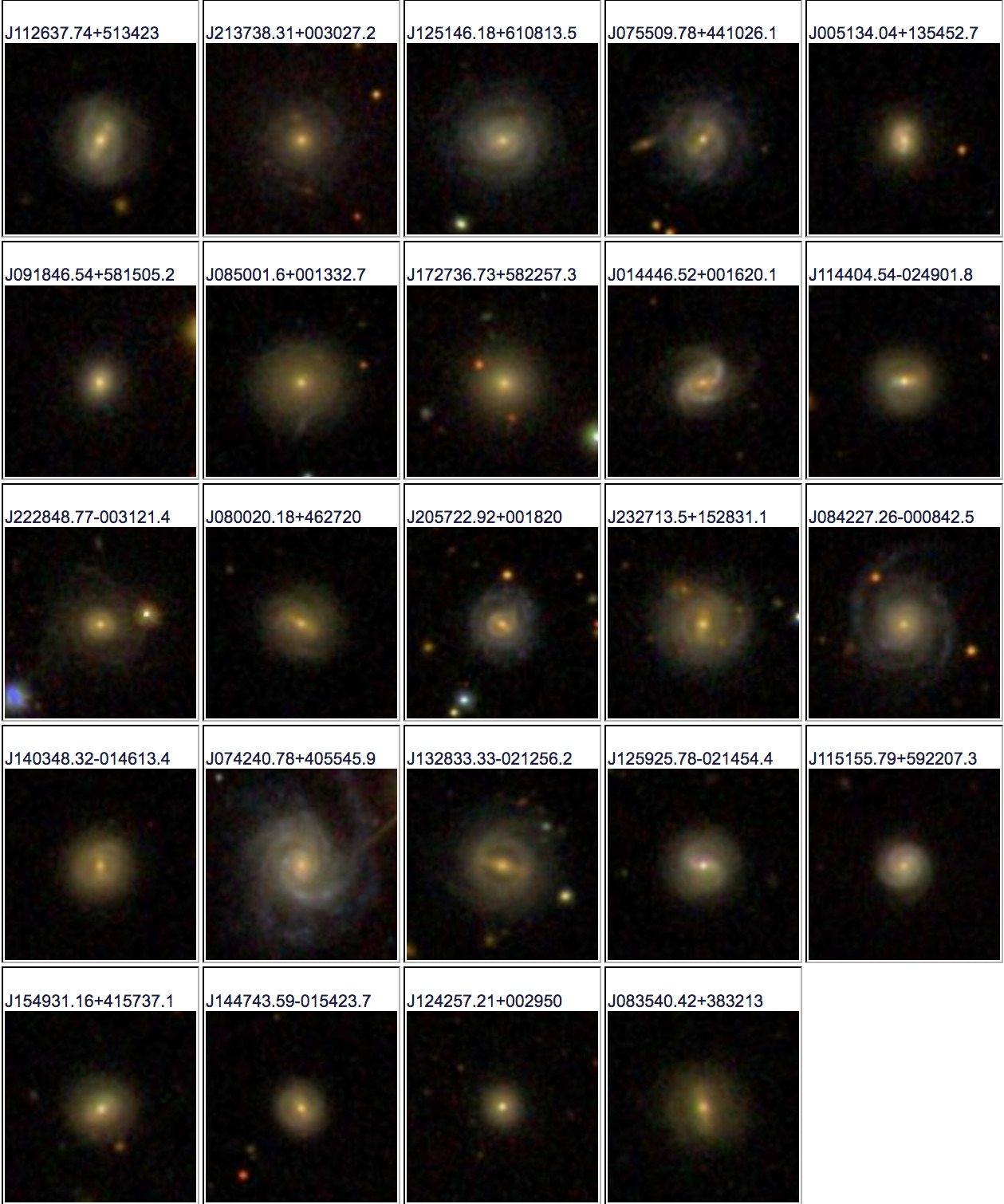}
\caption{Few galaxies from \citet{Gadotti09}'s sample that Random Forest classified as real bulges but \citet{Gadotti09} classified as pseudobulges.\label{fig:img_misPB}}
\end{figure*}

\section{RESULTS}\label{sec:res}
This section presents the pseudobulge frequencies of AGN and star-forming non-AGN galaxies, and the accretion properties of AGN in pseudobulges and real bulges. The AGN class includes both Seyferts and LINERs.
%The median, the 84th and 16th percentile deviations of the PDFs of the model parameters are given in the table

\subsection{Classifying bulge-types with Random Forest}

We classify $\sim 44,500$ face-on (axis ratio $b/a > 0.5$) SDSS galaxies above 10$^{10}$ M$_\odot$ and at $z < 0.07$ into real bulges or pseudobulges with $\sim 93 \pm 2 $\% training accuracy, and 96\% test accuracy using a Random Forest algorithm, remarkably with only 647 galaxies as our training sample \citep{Gadotti09}. The training accuracy is estimated by a ten fold cross-validation. To estimate the testing accuracy, 162 galaxies (20\% of \citet{Gadotti09}'s sample) are used after the algorithm is trained with 80\% of the sample. The Random Forest algorithm correctly classifies 124 galaxies as real bulges and 32 galaxies as pseudobulges. It misclassified 4 galaxies as real bulges and 2 as pseudobulges. Thus, the precision and recall for the real bulge sample are $\sim 97 \%$ and $\sim98$\% respectively while the true negative rate (for pseudobulge sample) is $\sim 89 \%$. Figure~\ref{fig:image_rf}, \ref{fig:img_misRB}, and \ref{fig:img_misPB} show example images of galaxies from \citet{Gadotti09}'s sample which are correctly classified or are misclassified by the Random Forest. We use structural and stellar population predictors that can easily be measured without resorting to careful bulge+bar+disc decomposition. With the advent of larger training samples based on high resolution galaxy images from future surveys, machine-based classification will likely be the most reliable and feasible approach to study bulge-types of large samples ($\gtrsim10^{4}$) of galaxies. 

\begin{figure*}
\includegraphics[width=0.48\textwidth]{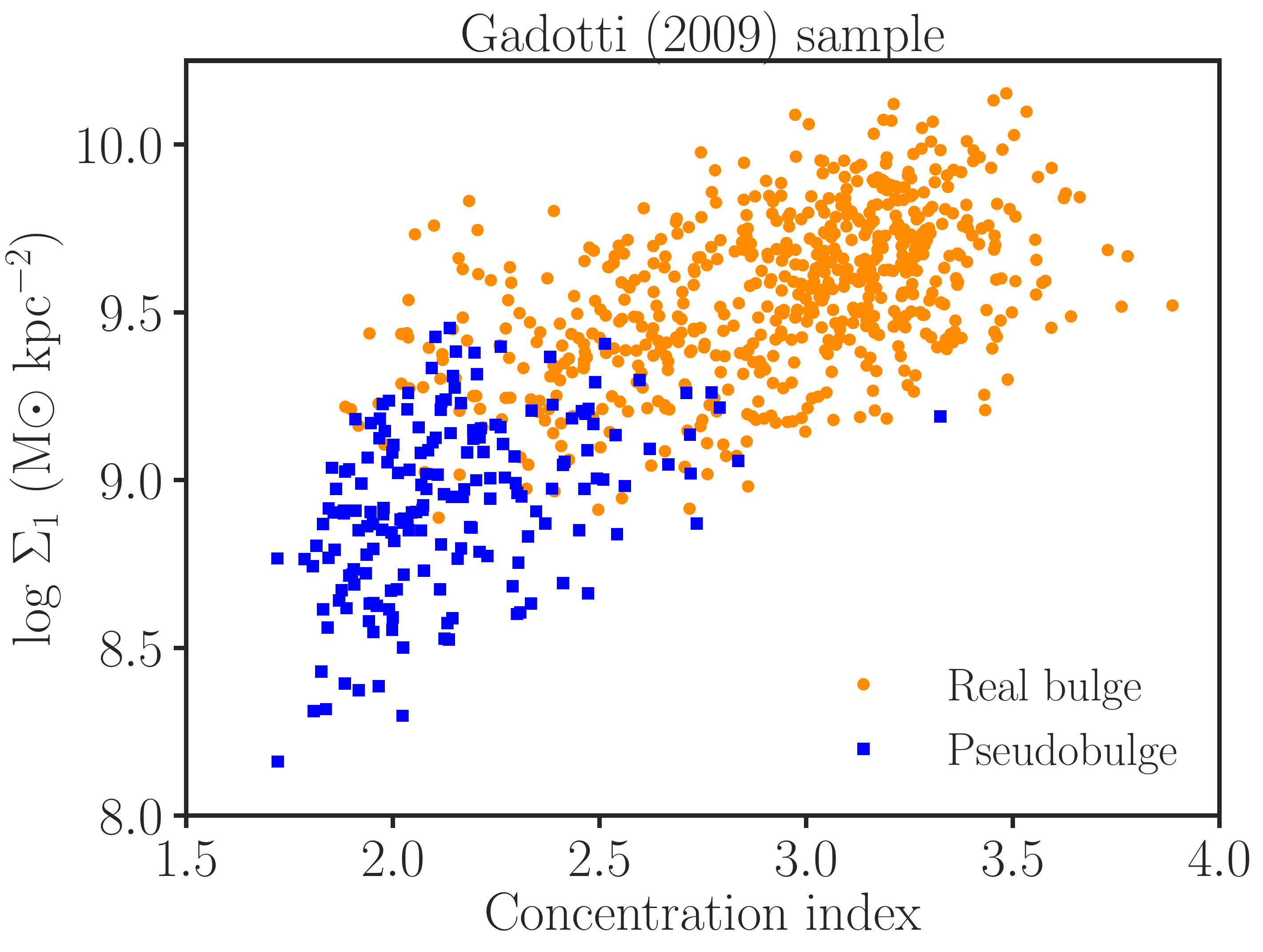}
\includegraphics[width=0.48\textwidth]{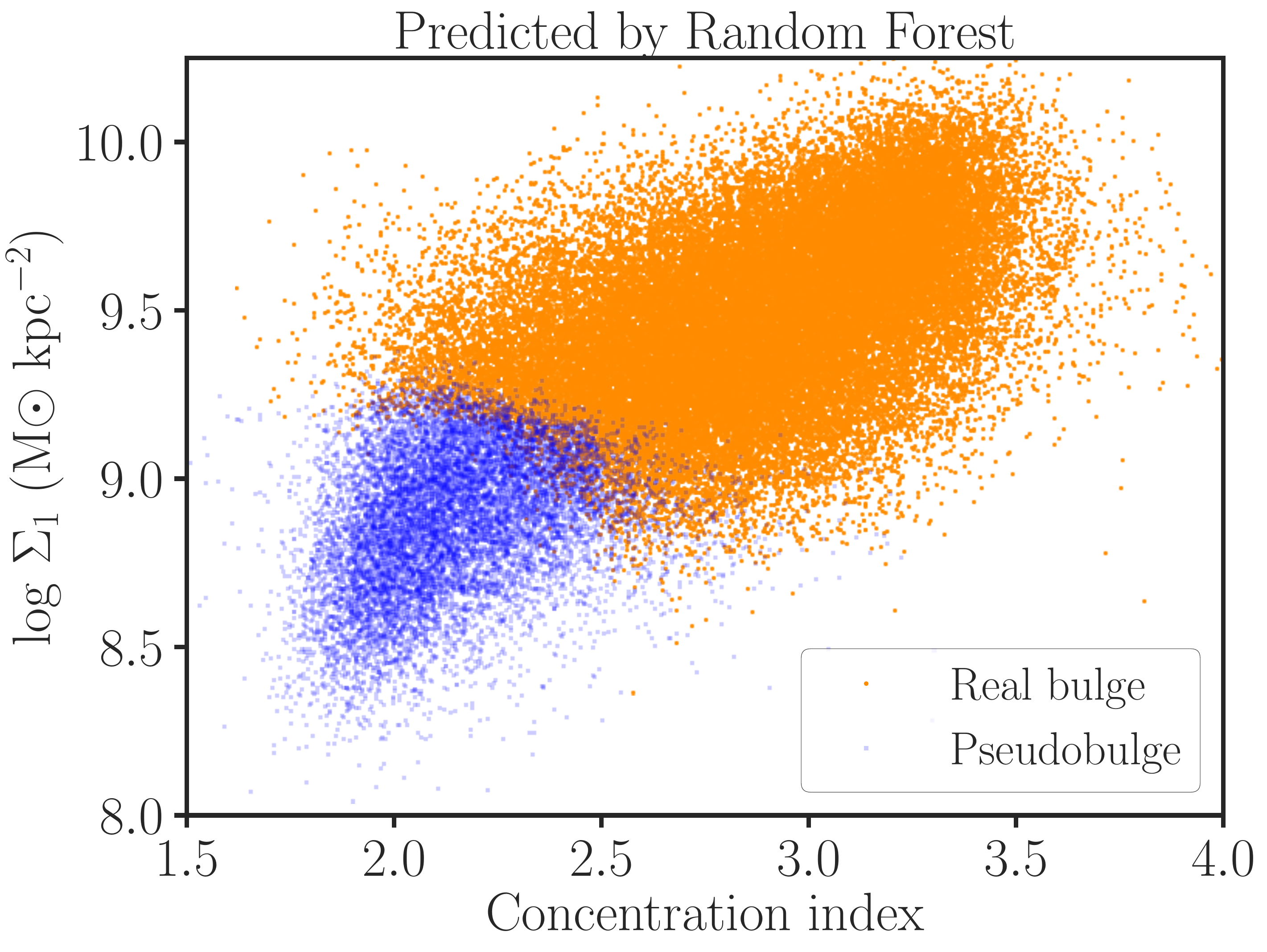}
\includegraphics[width=0.48\textwidth]{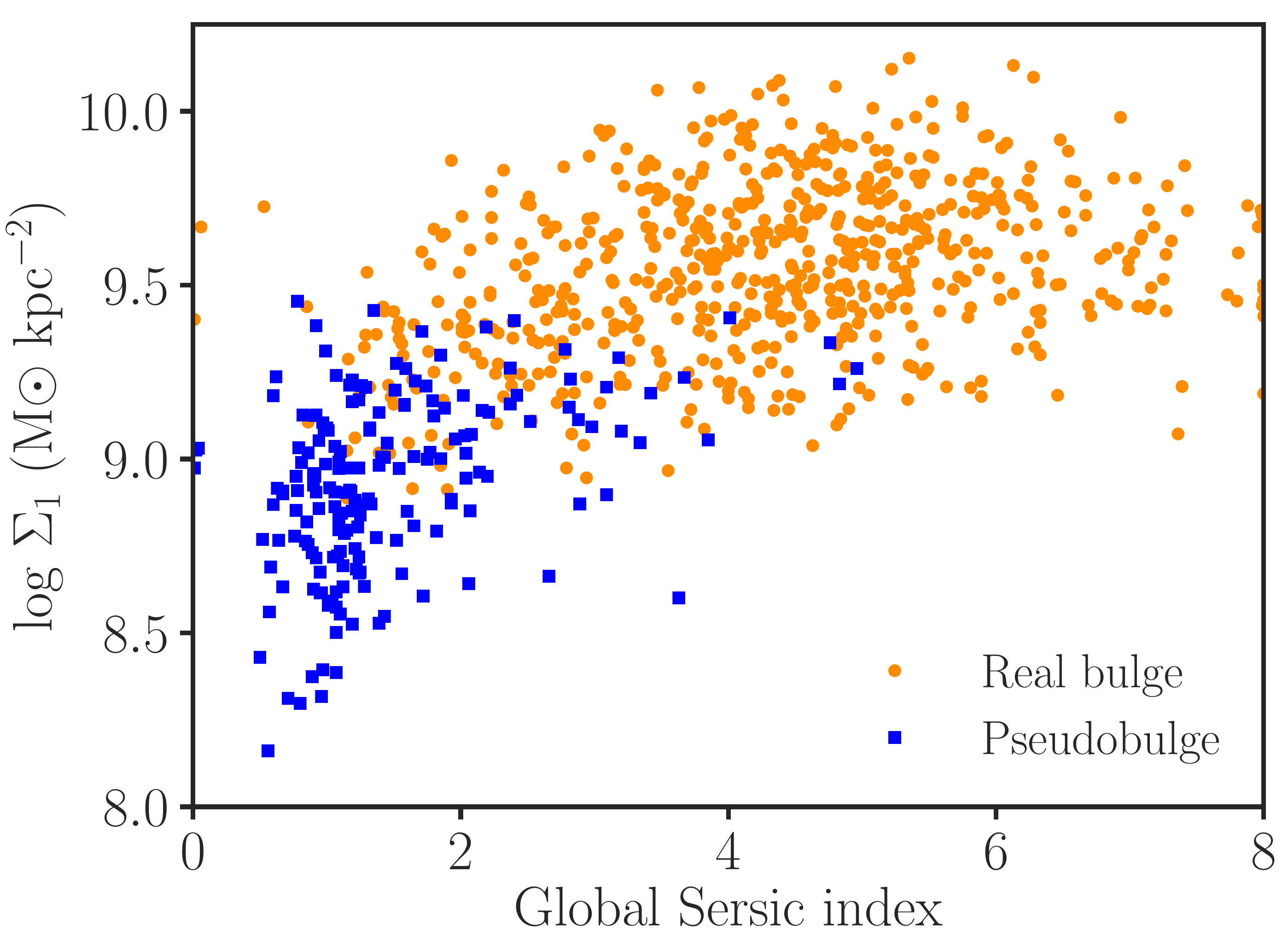}
\includegraphics[width=0.48\textwidth]{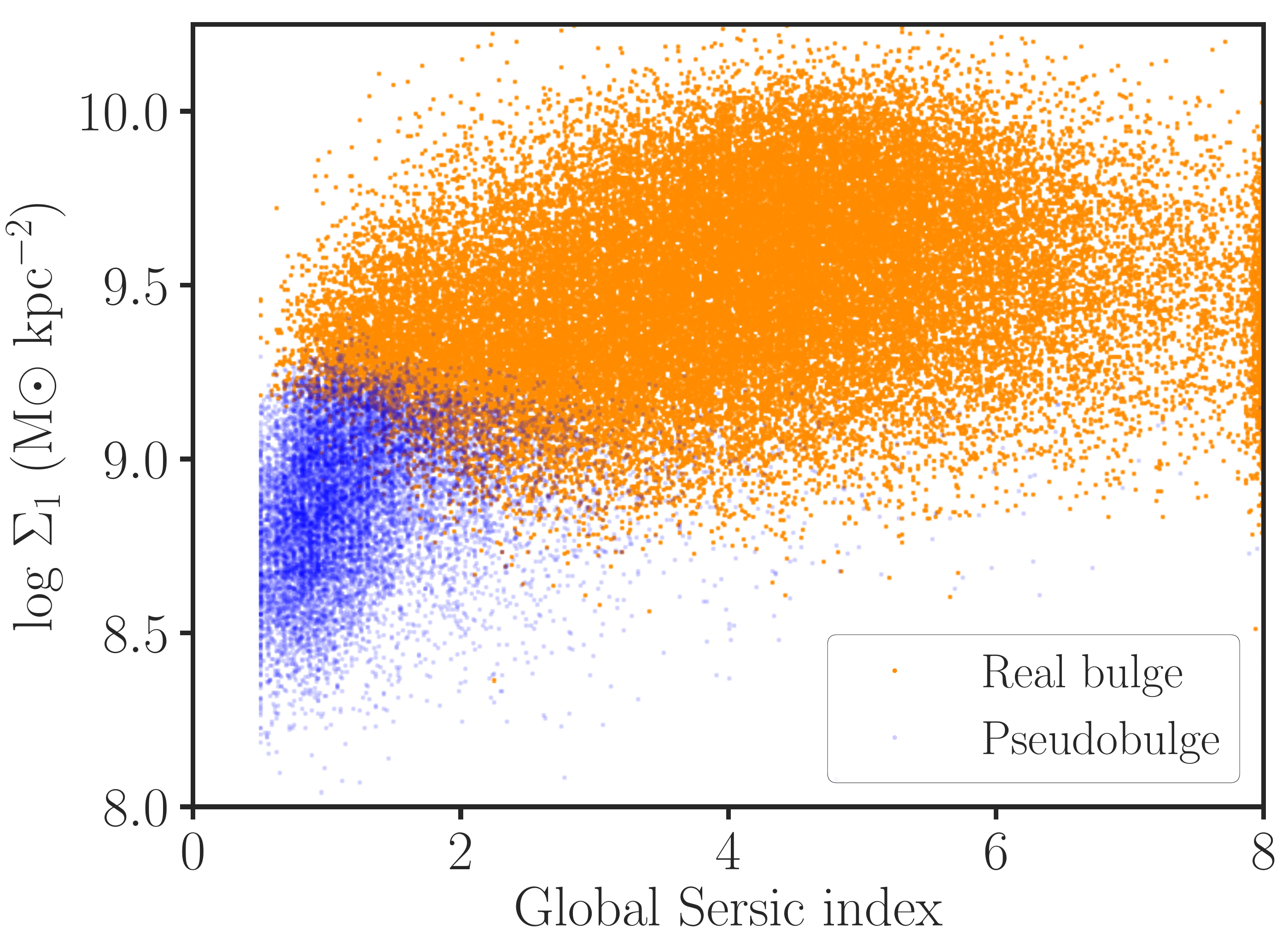}
\includegraphics[width=0.48\textwidth]{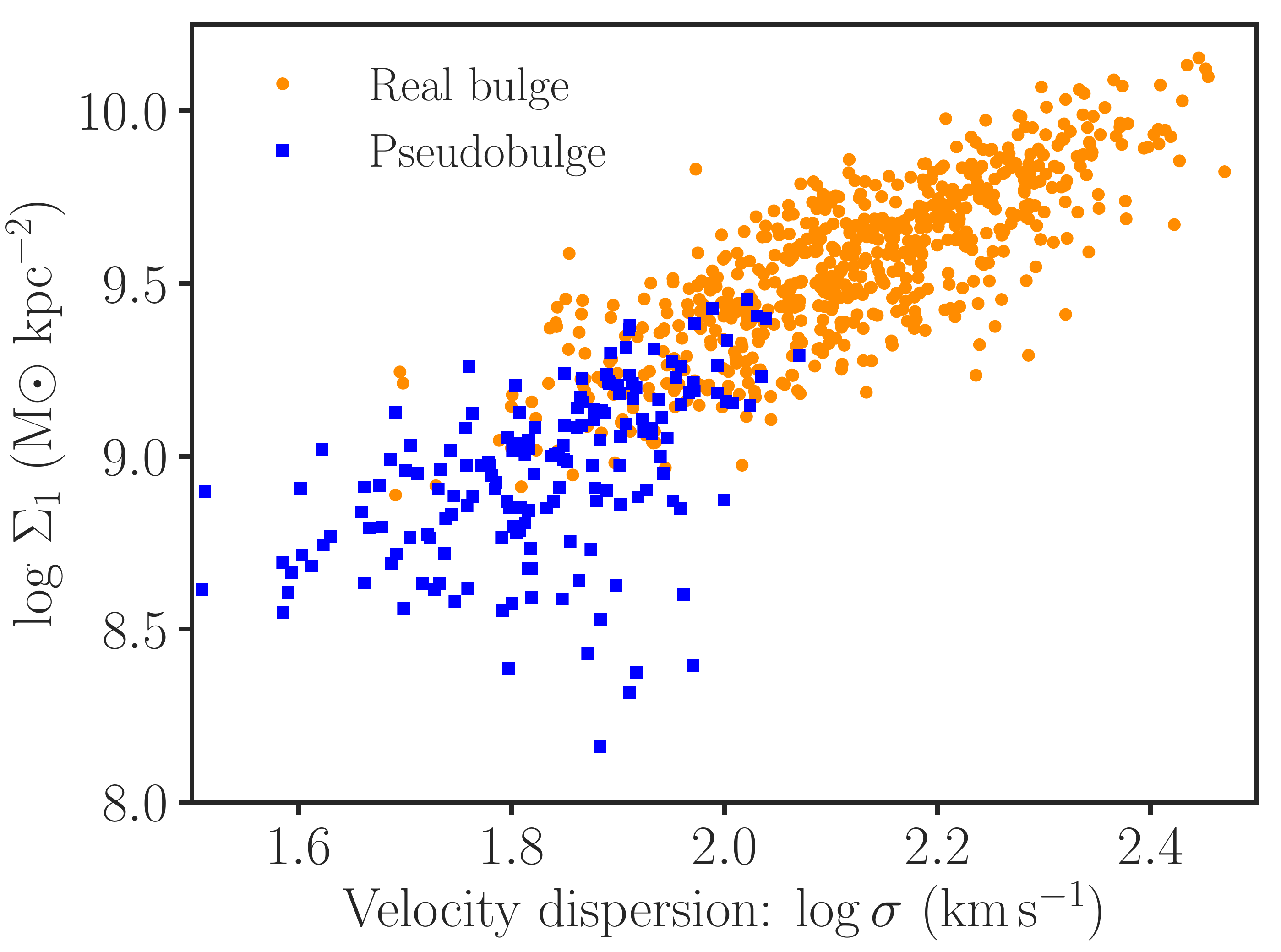}
\hfill
\includegraphics[width=0.48\textwidth]{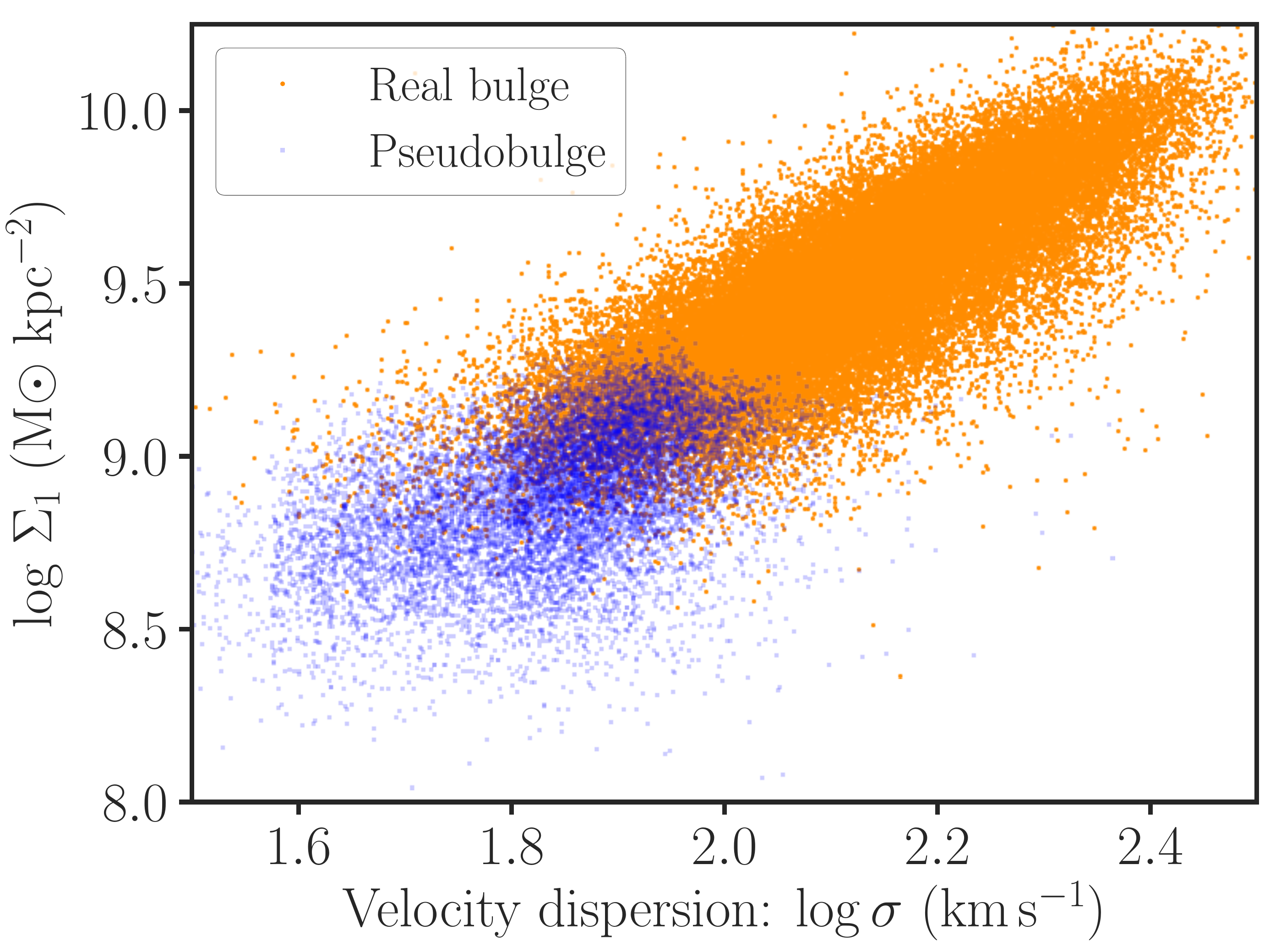}
\caption{The bulge-types of \citet{Gadotti09}'s sample (left), and the predicted types by Random Forest (RF, right). The central mass density within 1 kpc, $\Sigma_1$, the S\'{e}rsic index, and the concentration index are among important parameters suggested by the algorithm. The RF classification maps well the bulge-types from the training sample to the predicted sample. Note that the classification boundaries are not just vertical or horizontal lines, and that the two bulge-types overlap near the boundary in a given figure, indicating that combining multiple parameters is useful in cleanly separating the two bulge-types.\label{fig:train_pred}}
\end{figure*}

Figure~\ref{fig:train_pred} compares the bulge-types of \citet{Gadotti09}'s sample and our predicted types in the parameter spaces that the RF algorithm identifies as important. The central mass density within 1 kpc, $\Sigma_1$, the concentration index, and the global S\'{e}rsic index are among the important parameters for predicting bulge-types. The left panels show \citet{Gadotti09}'s sample and the right panels show the predicted bulge-types by the RF algorithm for a large SDSS sample. The blue points are pseudobulges and the orange points are real bulges. This distinction is based on Gadotti's classes, which are based on deviations from the Kormendy relation for bulges, as described in Section 1. The figure shows that the algorithm maps the bulge-types well from the training sample to the predicted sample.

\subsection{The weight of $\Sigma_1$ in predicting bulge-types}

%\textcolor{red}{Sandy and Yifei  please add your thoughts, and some discussion of Luo in prep.} \\

The central density within 1 kpc, $\Sigma_1$, is one of the best predictors of star formation quenching \citep[e.g.,][]{Fang+13, vanDokkum+14, Woo+15, Barro+17}, and may be linked to the properties of supermassive black holes (Chen et al. in prep.). Based on our analysis, there is a suggestive evidence that the central mass density within 1 kpc, $\Sigma_1$ is a useful quantity in predicting bulge-types (Figure~\ref{fig:train_pred}) and it may have comparable usefulness as the S\'{e}rsic and concentration indices \citep[see also][]{Luo+19}.

\begin{figure}
\includegraphics[width=0.47\textwidth]{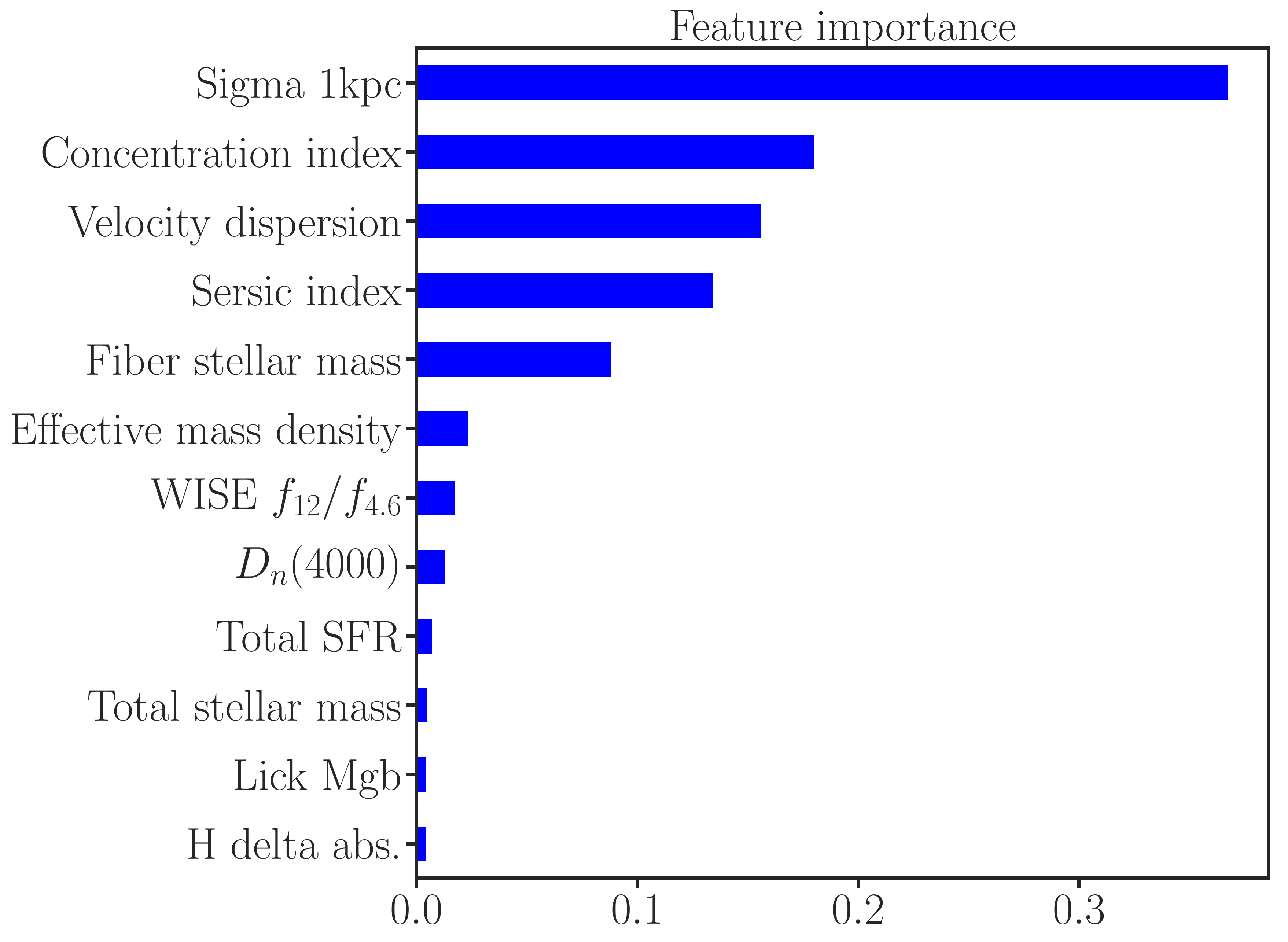}
\caption{The importance of galaxy properties, according to Random Forest, for predicting bulge-types defined on the basis of the Kormendy relation \citep{Gadotti09}. The RF algorithm hints that the central central mass density within 1 kpc, $\Sigma_1$, is at least as important as concentration and S\'{e}rsic indices in predicting bulge-types. However, the precise rank of these predictors is not well determined ($\sim1\sigma$) based on the current training data.\label{fig:imp}}
\end{figure}

Figure~\ref{fig:imp} shows the variable importance for predicting the two bulge classes, as defined by \citet{Gadotti09}. Our machine based classification has the advantage of combining multiple important parameters. The top three predictors in decreasing importance are the central mass density within 1 kpc, $\Sigma_1$, the concentration index and the global S\'{e}rsic index. Note that the importance of the variables is marginally constrained ($\sim1\sigma$). The error of the feature importance for a parameter can be computed from the standard deviation of its importance across all trees in the forest.  A larger training sample is needed to accurately rank the variables. When future high resolution data on nearby galaxies are available, the importance of $\Sigma_1$ in bulge-type classification should be revisited.

% The ranking also depends on how a pseudobulge is defined. If we divide the sample into real bulges and pseudobulges using $n_b$, the S\'{e}rsic index, $\Sigma_1$ and concentration index are the top three important predictors. \citet{Gadotti09} claimed that the concentration index is a better proxy for the bulge-to-total ratio than the global S\'{e}rsic index. 

\subsection{Comparing the pseudobulge fractions in AGNs star-forming galaxies}

\begin{figure*}
\includegraphics[width=0.48\textwidth]{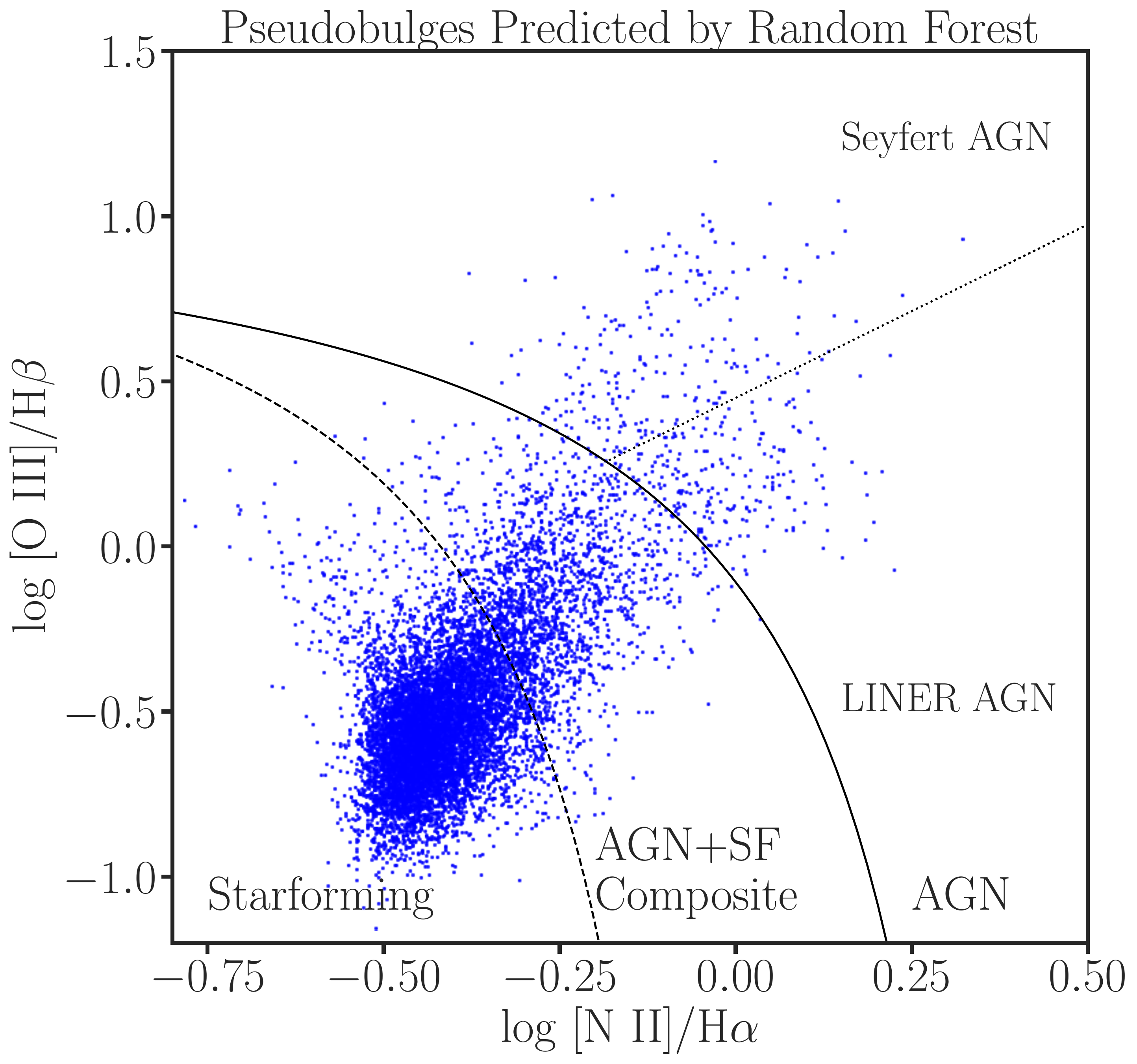}
\includegraphics[width=0.48\textwidth]{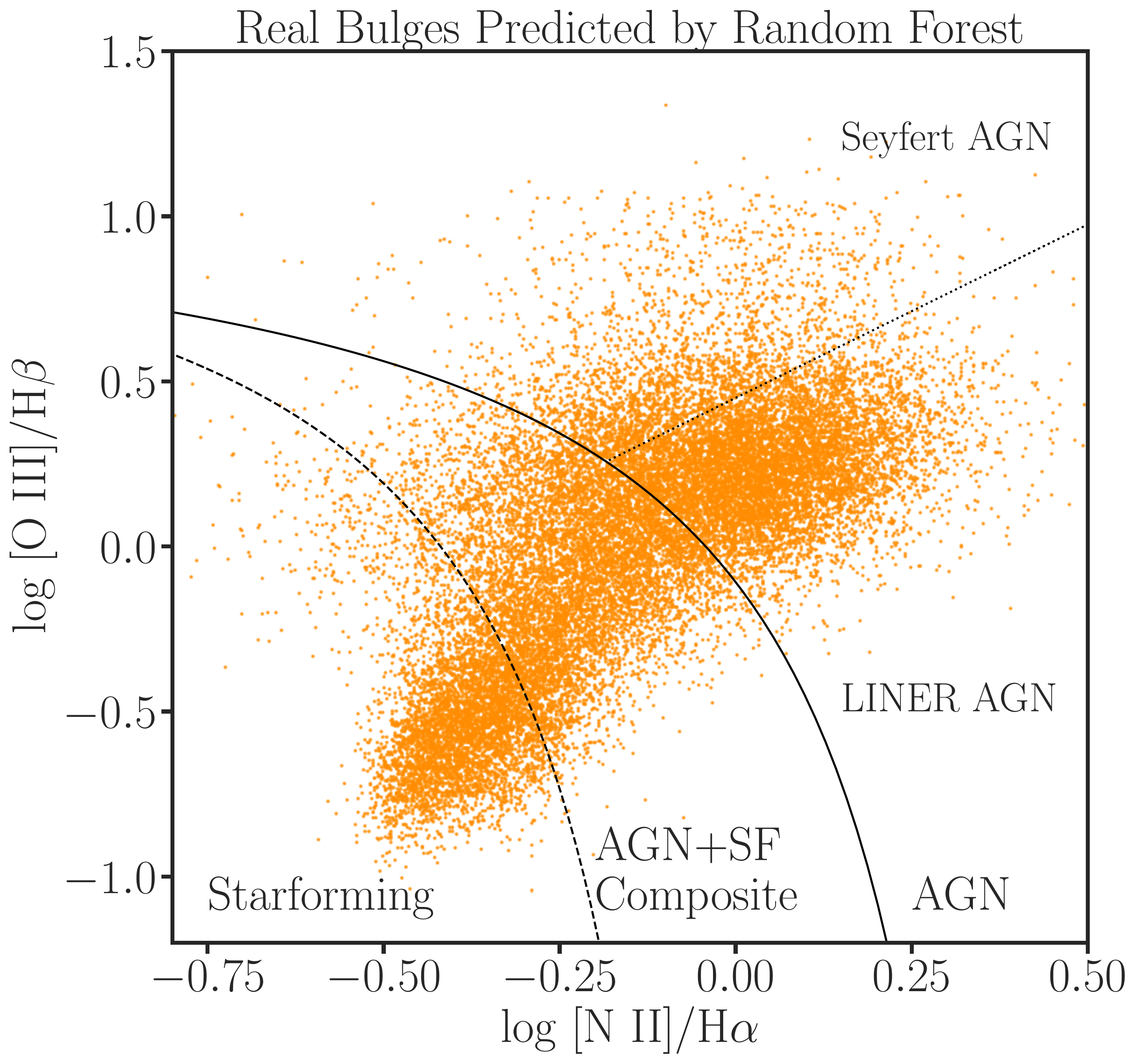}
\caption{Emission-line ratio AGN diagnostic for the pseudobulges (left) and real bulges (right) whose four emission lines have S/N $>2$. The solid curve denotes the theoretical boundary for strong starbursts \citep{Kewley+01}, while the dashed curve denotes the empirical boundary of pure star-forming galaxies \citep{Kauffmann+03}. Galaxies between the two curves have both star formation and AGN. The dotted line in the AGN region splits the AGNs into Seyferts and LINERs \citep{Schawinski+07}. The fraction of galaxies with real bulges is higher when they host AGN, indicating a link between AGN triggering and bulge growth. \label{fig:bpt}}
\end{figure*}

Figure~\ref{fig:bpt} shows the [\ion{O}{3}]/{\rm H}$\beta$ versus [\ion{N}{2}]/{\rm H}$\alpha$ line ratios \citep{Baldwin+81}. This line ratio diagram (also known as the BPT diagram) distinguishes between emission lines from \ion{H}{2} regions and AGNs. AGN-dominated galaxies have larger ratios in both axes and occupy the upper right part of the diagram, while star-forming galaxies occupy the lower left. The solid black curve demarcates the theoretical boundary for extreme starbursts, and galaxies above the curve are inconsistent with starbursts and likely host AGNs \citep{Kewley+01}. The dashed curve demarcates the empirical lower boundary for AGNs \citep{Kauffmann+03}. Objects below this curve are likely star-forming galaxies without AGNs. Galaxies between the two curves are mostly composites of star formation and AGN. The blue points in the left panel of Figure~\ref{fig:bpt} are pseudobulges predicted by the Random Forest algorithm, while the orange points in the right panel are real bulges (classical bulges or elliptical galaxies). All four emission lines of a galaxy are required to be $>2\sigma$ for a galaxy to be plotted on the figure. %Since elliptical galaxies have very weak emission almost all galaxies in the panel are classical bulges (this can also be shown by predicting classical bulges and ellipticals separately). 

\begin{figure}
\includegraphics[width=0.49\textwidth]{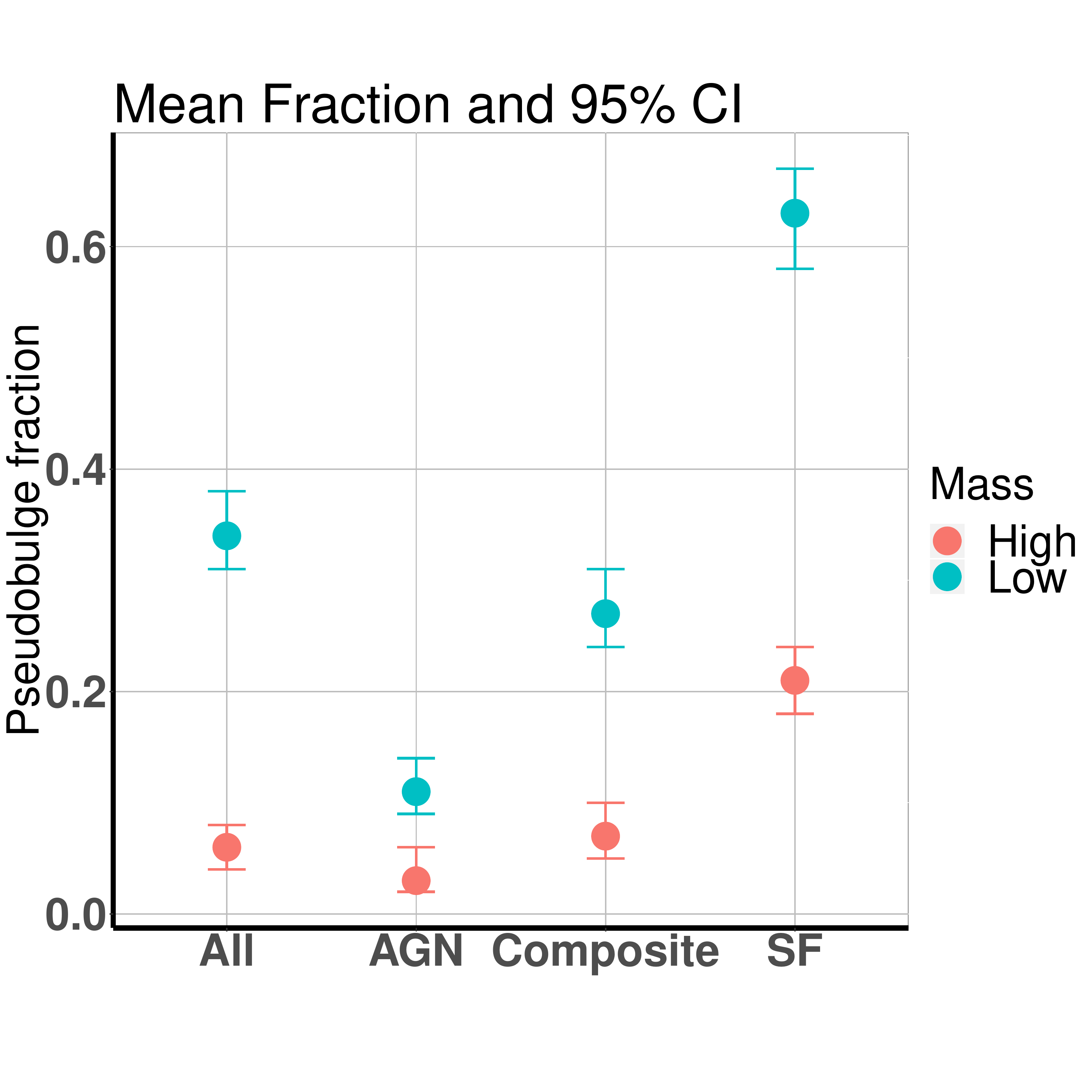}
\includegraphics[width=0.49\textwidth]{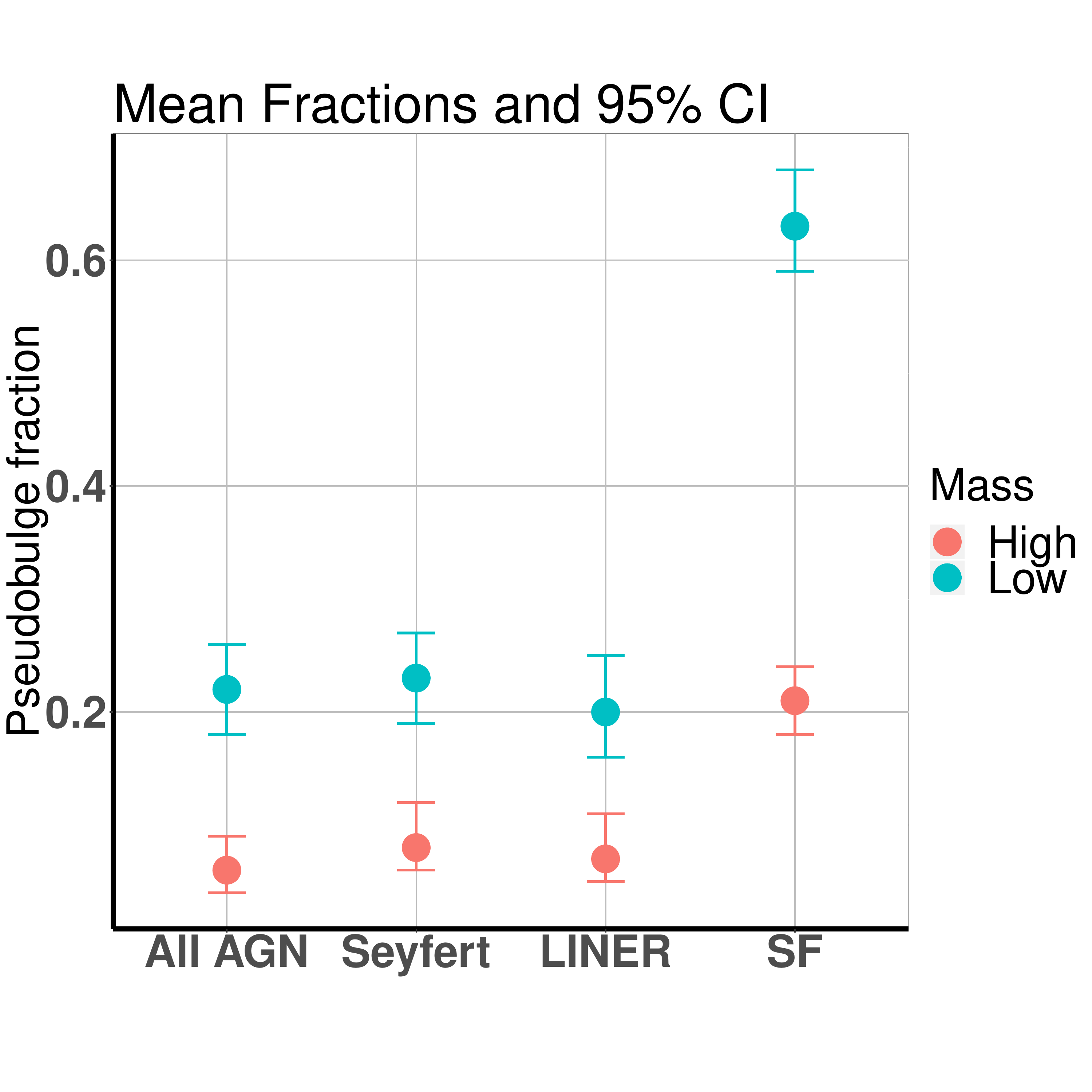}
\caption{The mean pseudobulge fractions and the 95\% confidence intervals for a large sample of SDSS galaxies. The sample is split into low ($\rm{\log M \,(M_\odot) < 10.5}$, cyan), and high stellar masses ($\rm{\log M \,(M_\odot) > 10.5}$, red)  or/and into different regions of the BPT emission-line ratio diagram as shown in Figure~\ref{fig:bpt}. The galaxies plotted on the bottom panel are restricted to have H$\alpha \ > 3${\AA} to minimize contamination from LINER-like fake AGNs \citep{CidFernandes+11}. \label{fig:pp}}
\end{figure}

We compute the pseudobulge fraction in the star-forming, AGN+SF composite, and AGN regions of the line ratio diagram as the number of pseudobulges in a given region divided the total number of galaxies in that region. Table~\ref{tab:gad_frac} and Figure~\ref{fig:pp} present the mean fractions and the 95\% confidence intervals (CI) in different regions. For the total sample shown in Figure~\ref{fig:bpt}, the pseudobulge fractions significantly decrease from 54\% (CI: 50--58\%) in the star-forming region to 18\% (CI: 15--21\%) in the composite region and to 5\% (CI: 3--8\%) in the AGN region. Since the pseudobulge fraction is known to depend strongly on stellar mass \citep{Fisher+10}, we divide the sample into low and high mass using $\rm{\log M \,(M_\odot) = 10.5}$ as a threshold. The pseudobulge fractions are significantly lower in AGNs than in star-forming galaxies both at high and low masses. At $\rm{\log M \,(M_\odot) =10 - 10.5}$, the fraction decreases from  63\% (CI: 58--67\%) in star-forming galaxies to 11\% (CI: 9--14\%) in AGNs. Likewise, at $\rm{\log M \,(M_\odot) = 10.5-12}$, it decreases from 21\% (CI: 18--24\%) to 3\% (CI: 2--6\%). Composite galaxies also have significantly lower pseudobulge fractions than do star-forming galaxies in the two mass ranges. 

Low ionization emission line regions (LINERs) may have a stellar origin, and some of them may not be weak AGNs, especially if their H$\alpha$ equivalent width is less than 3{\AA} \citep{CidFernandes+11}. We, thus, further divide our AGN sample into LINERs and Seyferts using the dotted line in Figure~\ref{fig:bpt} \citep{Schawinski+07}. Table~\ref{tab:gad_frac2} presents pseudobulge fractions after restricting the AGN and star-forming samples to have H$\alpha \ >3${\AA}. The pseudobulge fractions are still lower ($\sim 3 \times$) in both Seyferts and LINERs than in star-forming galaxies after the H$\alpha$ cut.

\begin{deluxetable*}{lccccc}
\tabletypesize{\footnotesize}
\tablecolumns{6} 
\tablewidth{0pt} 
\tablecaption{The pseudobulge fraction of AGN and star-forming galaxies for bulge-types defined according to \citet{Gadotti09}. The AGN class includes both Seyferts and LINERs.\label{tab:gad_frac}}
\tablehead{
		\colhead{Mass} & \colhead{All AGN} & \colhead{Seyfert  AGN} & \colhead{AGN+SF Composite} & \colhead{Star-forming (SF)} &  \colhead{All}}
	\startdata
		$\log M = 10 - 10.5 $ & 0.11\ (0.09, 0.14) & 0.14\ (0.11, 0.17) & 0.27\ (0.24, 0.31) & 0.63\ (0.58, 0.67) & 0.34\ (0.31, 0.38) \\ 
	         $\log M = 10.5-12 $ & 0.03\ (0.02, 0.06) & 0.06\ (0.04, 0.09) & 0.07\ (0.05, 0.10) &  0.21\ (0.18, 0.24) & 0.06\ (0.04, 0.08)  \\ 
		All & 0.05\ (0.03, 0.08) & 0.09\ (0.07, 0.12) & 0.18\ (0.15, 0.21) & 0.54\ (0.50, 0.58) & 0.21\ (0.18, 0.24) \\
		\enddata
		\tablecomments{The numbers in the parentheses are the 95\% confidence intervals.}
\end{deluxetable*}

Furthermore, the pseudobulge fractions computed above do not change significantly if we instead require the four emission lines to be detected at $>3\sigma$ or $>1\sigma$ or if we remove the signal-to-noise cut. Generally, real bulges have low SFR and weak AGN; their emission lines have lower signal-to-noise ratios than those of pseudobulges. For example, unlike real bulges, most pseudobulges have signal-to-noise ratios of H$\beta$ emission-line greater than 5. The signal-to-noise distributions of the [\ion{O}{3}] emission-line for the two bulge-types are similar. Less stringent signal-to-noise cut increases the AGN fraction in real bulges and helps strengthen the conclusion that AGNs are more common in real bulges than in pseudobulges. The fact that the pseudobulge fraction is only $\sim 20$\% in the composite region of the BPT diagram also suggests that the dilution of the AGN signature by the star formation in pseudobulges is not a significant effect to change this conclusion.

Moreover, we split our sample into high and low redshifts bins and compute the pseudobulge fractions in AGN, composite and star-forming galaxies. The results of this analysis are presented in Table~\ref{tab:z_pfrac}. We observe a noticeable decrease in the pseudobulge fractions in AGN and composite galaxies of $\sim 5\%$ as the fiber diameter increases from $\sim 1.2 - 1.8$ kpc ($z = 0.02 - 0.03$) to $\sim 3.5 - 4.0$ kpc ($z = 0.06-0.07$). Perhaps this change is because, with increasing redshift, the SDSS fiber encloses more circum-nuclear area, and star formation from this area dilutes the AGN signature. The fiber effect, nevertheless, does not change our conclusion that a large majority ($\gtrsim 75\%$) of AGNs are hosted by galaxies with real bulges. %Having repeated similar analyses,  we also do not find evidence that suggests that our other main conclusions originate from the observational bias due the fiber aperture effect. It will be valuable to confirm or refute our results with high spatial resolution data.

Recently, \citet{Agostino+19} assessed the reliability of the BPT diagram to identify galaxies hosting AGN by using 332 X-ray AGNs that have all four BPT emission lines detected at $>2\sigma$. Only 34 out of the 332 X-ray AGNs were found within the star-forming region of the BPT diagram. The authors did not find that the star formation dilution can satisfactorily explain the apparent misclassification of these X-ray AGNs. On the other hand, the BPT diagram cannot classify about 40\% of all X-ray AGNs with very weak emission lines and do not satisfy the signal to noise requirement. The authors pointed out that these galaxies tend to also have very low specific star formation rates. According to the authors, the most likely explanation for the X-ray AGNs found in the star-forming region of the BPT diagram is that they have intrinsically weak AGN lines, and are only classifiable by the BPT diagram when they have high SSFRs. If this is the case, the BPT is likely to misclassify or unable to classify preferentially real bulges more than pseudobulges as the latter tend to have higher SSFRs and accretion luminosities (more on this point in the next section). The dominant bulge-type in AGNs and composite galaxies is a real bulge, and adding 10\% to the pseudobulge fraction, for the failure of the BPT diagram when it can be used, does not explain away the strong preference of AGNs for real bulge hosts. 

\begin{deluxetable*}{lccccc}
\tabletypesize{\footnotesize}
\tablecolumns{6} 
\tablewidth{0pt} 
\tablecaption{The pseudobulge fraction with H$\alpha$ equivalent width cut for bulge-types defined according to \citet{Gadotti09}. \label{tab:gad_frac2}}
\tablehead{
\colhead{Mass} & \colhead{All AGN (H$\alpha > 3${\AA})} & \colhead{Seyfert (H$\alpha > 3${\AA})} & \colhead{Seyfert (H$\alpha > 6${\AA})} & \colhead{LINER (H$\alpha > 3${\AA})} & \colhead{SF (H$\alpha > 3${\AA})}} %\\
\startdata
       $\log M = 10 - 10.5 $ & 0.22\ (0.18, 0.26) & 0.23\ (0.19, 0.27) & 0.23\ (0.19, 0.28) & 0.20\ (0.16, 0.25) & 0.63\ (0.59, 0.68) \\ 
       $\log M = 10.5-12 $ & 0.06\ (0.04, 0.09) & 0.08\ (0.06, 0.12) & 0.09\ (0.06, 0.13) & 0.07\ (0.05, 0.11) & 0.21\ (0.18, 0.24) \\ 
       All & 0.12\,(0.10, 0.15) & 0.15\ (0.12, 0.18) & 0.15\ (0.12, 0.19) & 0.10\ (0.08, 0.13) & 0.55\ (0.51, 0.59) \\
\enddata
 \tablecomments{According to \citet{CidFernandes+11}, photoionization by old stellar populations may explain weak H$\alpha < 3${\AA} emission without invoking AGN. They consider an AGN to be a Seyfert (a strong AGN) if it has H$\alpha > 6${\AA}.}
\end{deluxetable*}

\begin{figure*}
\includegraphics[width=0.48\textwidth]{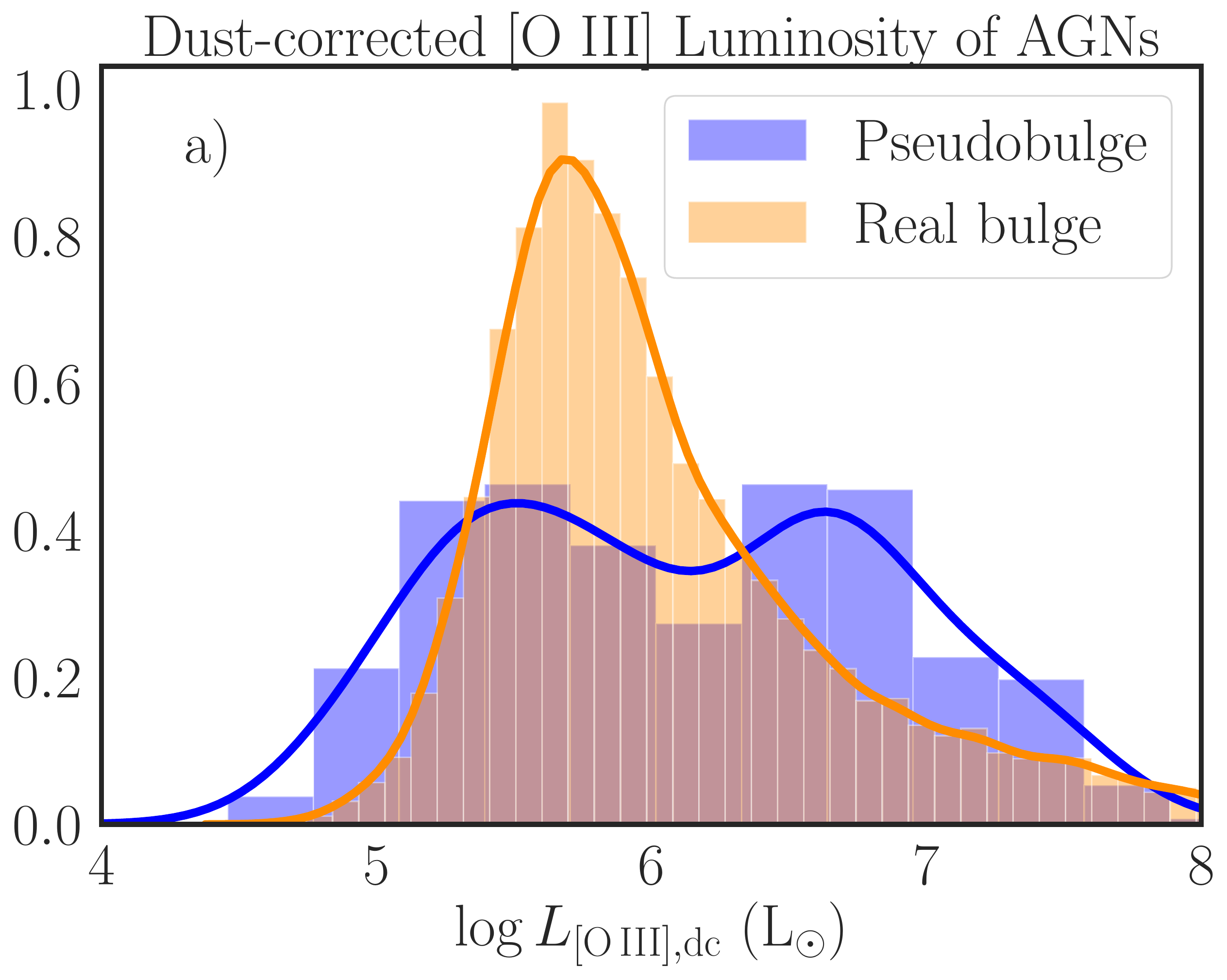}
\includegraphics[width=0.48\textwidth]{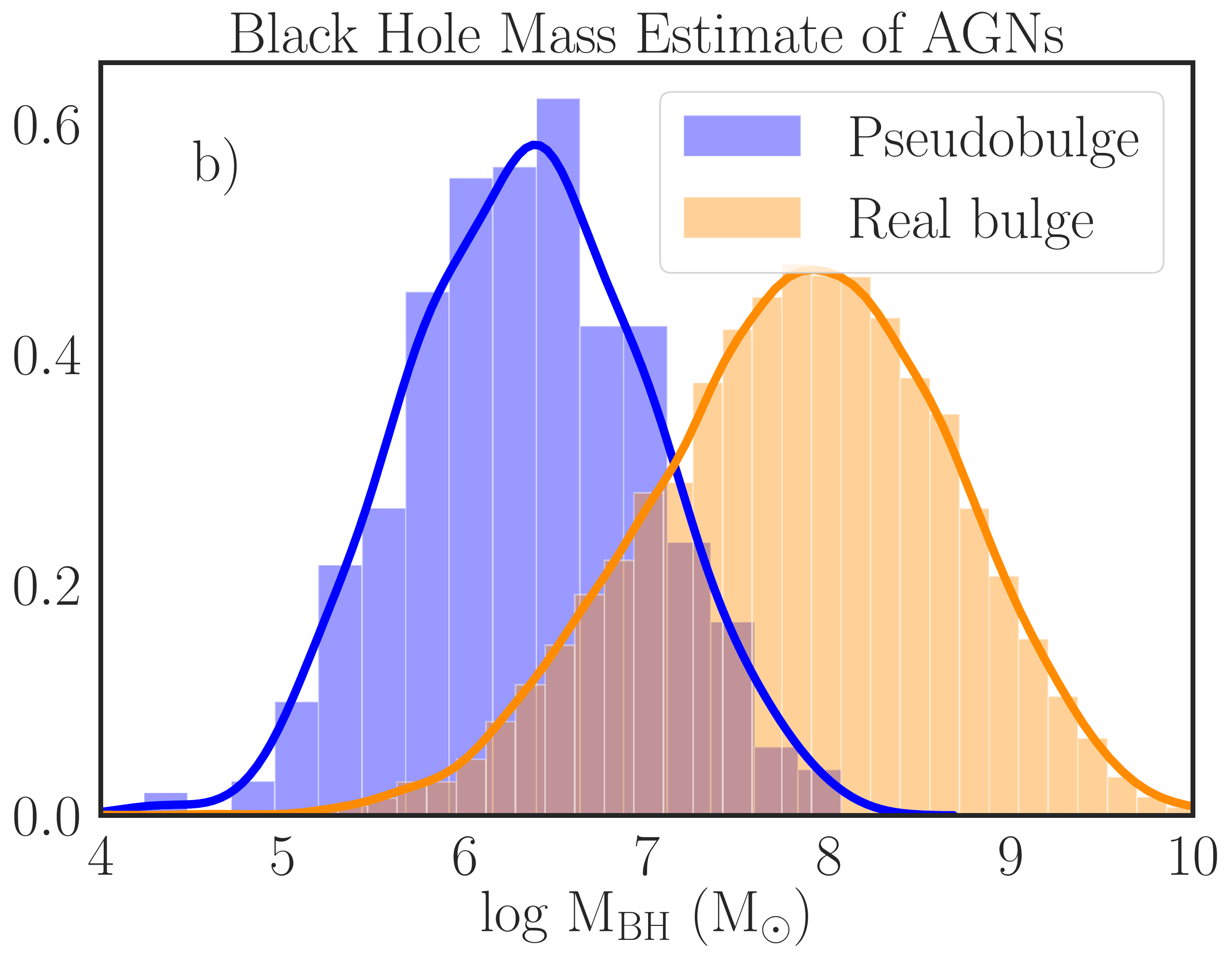}
\includegraphics[width=0.48\textwidth]{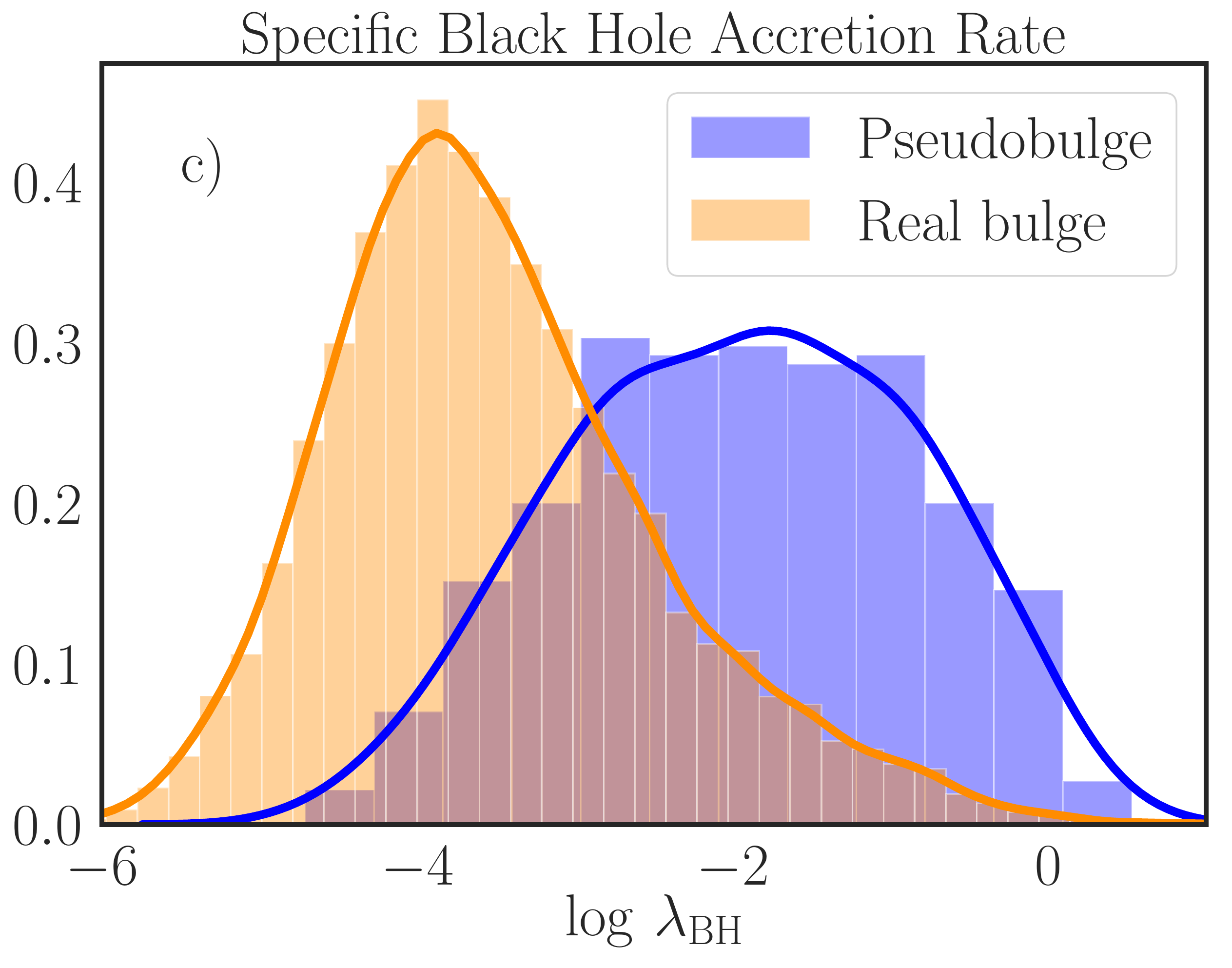}
\hfill
\includegraphics[width=0.48\textwidth]{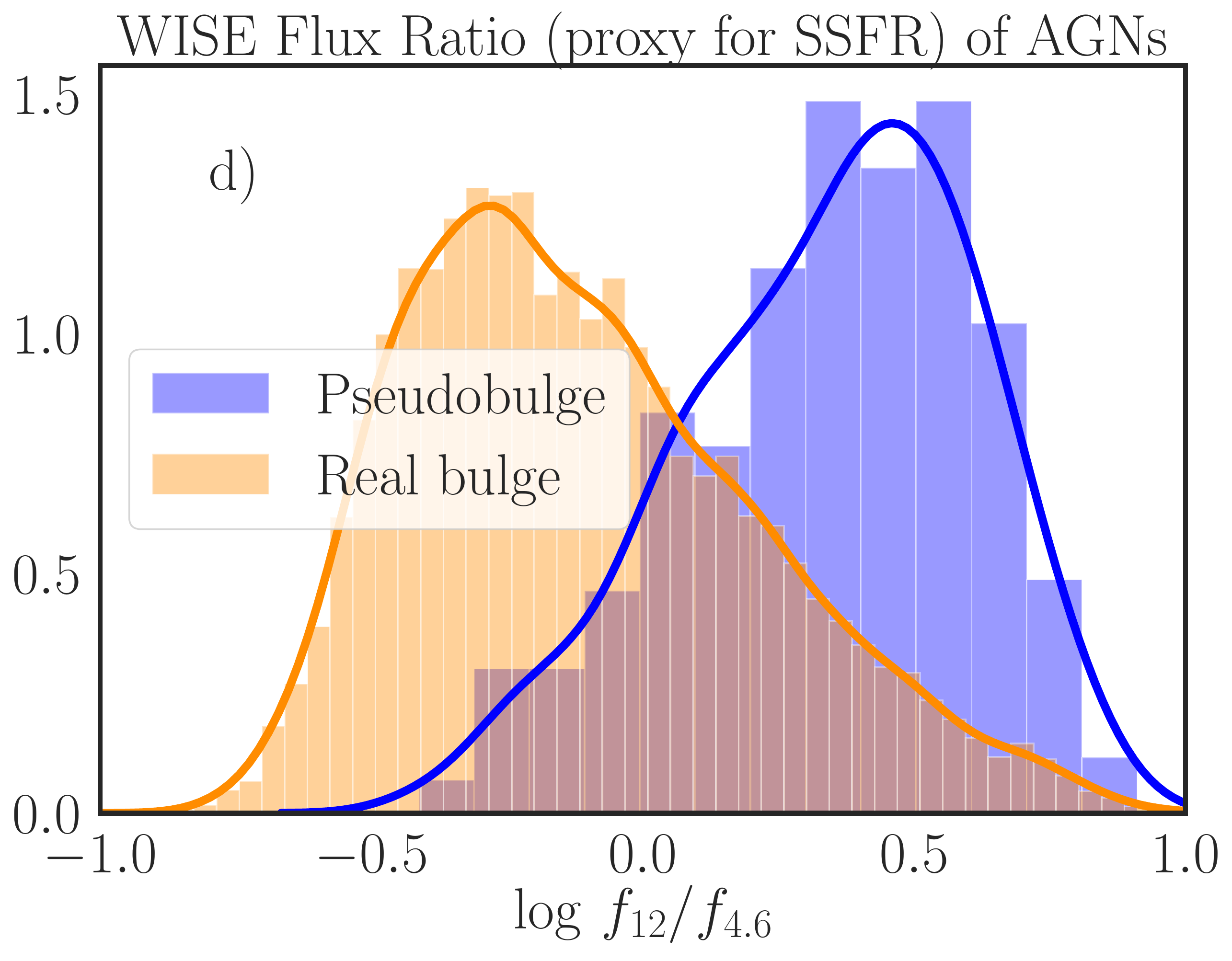}
\caption{The normalized distributions of dust-corrected \ion{O}{3} luminosities, black hole masses, Eddington ratios ($\propto$ \ion{O}{3} luminosity per black hole mass), and WISE 12\,$\mu$m flux to  4.6\,$\mu$m flux ratio ($\propto$ gas fraction and specific star formation rate) for all AGNs in real bulge or pseudobulge host galaxies. All distributions are significantly different for the two bulge-types. The fits are kernel density estimates, done using \texttt{Seaborn} package. \label{fig:o3dist}}
\end{figure*}

\subsection{Comparing accretion properties of black holes in real bulges and pseudobulges}

Figure~\ref{fig:o3dist}a shows the distributions of dust-corrected [\ion{O}{3}] 5007{\AA} luminosities ($\log \, L_{{\rm O3,dc}}$[L$_\odot]$) for AGNs which are pseudobulges (blue) and real bulges (orange). The [\ion{O}{3}] 5007{\AA} luminosity is known to correlate with the hard X-ray luminosity and is a good proxy for the accretion luminosity \citep{Heckman+04}. The Kolmogorov-Smirnov (KS) test shows that the null hypothesis that the [\ion{O}{3}] distributions of pseudobulges and real bulges are the same can be rejected at $>5\sigma$ (the KS statistic, D, the maximum difference, is 0.2 and $p$ value $\approx$ 0). The 16, 50, and 84 percentiles of $\log \ L_{{\rm O3,dc}}$ of pseudobulge AGNs are 5.3, 6.1, and 6.9 respectively. The corresponding percentiles for AGNs with real bulges are 5.5, 5.9, and 6.6. The [\ion{O}{3}] distributions are also different at $>5\sigma$ level before the dust-correction for the two bulge-types ($D = 0.2$, $p$ value $\approx 0$).

Figure~\ref{fig:o3dist}b displays the distributions of black hole mass estimates ($\log M_{\rm BH}$[M$_\odot$]). The black hole masses are estimated from stellar masses ($M_\star$) and half-mass radii $(R_{1/2})$ of the host galaxies using the black hole mass fundamental plane relation \citep{Bosch+16}, $\log M_{\rm BH} [M_\odot]= 7.48+2.91 \log \left( \frac{M_\star}{10^{11} {\rm M}_\odot}\right) - 2.77  \log \left( \frac{R_{1/2}}{5 \mathrm{kpc}}\right)$. The distribution of $\log M_{\rm BH} ({\rm M}_\odot)$ of pseudobulge AGNs are significantly shifted toward lower black hole masses compared to that of real bulges. The 16, 50, 84 percentiles of $\log \, M_{\rm BH}$ for AGNs with pseudobulges are 5.7, 6.4 and 7.0 respectively while those with real bulges are 7.0, 7.9, and 8.7 respectively. KS test indicates that the black hole mass estimates of the two bulge-types are significantly different ($D = 0.7$, $p$ value $\approx$ 0). %They also are different for the subsamples in the two mass ranges.

Figure~\ref{fig:o3dist}c shows the specific black hole accretion, $\log \lambda_\mathrm{BH}$, which is the bolometric luminosity divided by the Eddington luminosity, $L_\mathrm{Edd}$. Thus, estimating the bolometric luminosity from the [\ion{O}{3}] luminosity with a bolometric correction of $BC = 600$ \citep{Kauffmann+09} gives $\log \lambda_\mathrm{BH} = \log L_{{\rm O3}} + \log BC  - \log L_\mathrm {Edd} = \log L_{{\rm O3}} + \log BC - \log M_{\rm BH} - \log (3.2 \times 10^4)$. The average $\log \lambda_\mathrm{BH}$ of AGNs in real bulges is much smaller than those of AGNs in pseudobulges. The 16, 50, 84 percentiles of $\log \lambda_\mathrm{BH}$ for the pseudobulges are -3.1, -1.9 and  -0.8 respectively while they are -4.5, -3.6, and -2.5 respectively for real bulge AGN hosts. The $\log \lambda_\mathrm{BH}$ distributions for the two population are significantly different ($>5\sigma$) according to the KS test ($D = 0.5$ and $p$ value $\approx$ 0). 

It is known that the $\log L_{{\rm O3}}/M_{\rm BH}$ is correlated with specific star formation rate \citep[e.g.,][]{Kauffmann+07, Netzer09}. More star-forming and gas rich galaxies have more black hole accretion. Figure~\ref{fig:o3dist}d depicts the WISE 12\,$\mu$m flux to  4.6\,$\mu$m flux ratio, $\log f_{12}/f_{4.6}$, distribution. This ratio is a good proxy for SSFR in both star-forming galaxies and AGNs \citep{Donoso+12}. It also correlates strongly with molecular gas contents of galaxies \citep{Yesuf+17}. The distribution of $\log f_{12}/f_{4.6}$ for AGNs in real bulges is shifted toward lower values (lower SSFR and gas fraction) than that of AGNs in pseudobulges. The 16, 50, and 84 percentiles of $\log f_{12}/f_{4.6}$ for AGNs in real bulges are -0.44, -0.16 and 0.23 respectively while those in pseudobulges are 0.06, 0.38, and 0.61. These numbers may be converted to mean molecular gas to stellar mass ratios assuming the relation $\log f_{ \rm H_2}=1.13 \log f_{12}/f_{4.6}-1.8$ for all galaxies in the COLD GASS survey \citep{Yesuf+17,Saintonge+11}. For example, $\log f_{12}/f_{4.6} = -0.16$ corresponds to $f_{ \rm H_2}$ of $\sim 0.01$ and $\log f_{12}/f_{4.6} = 0.38$ corresponds to $f_{ \rm H_2}$ of $\sim 0.04$. The distributions presented in Figure~\ref{fig:o3dist} for the two bulge-types are also significantly different if we restrict the AGN sample to Seyfert AGNs only.

%The total gas fraction is about $4f_{\rm H_2}$. The Pearson correlation coefficient between $\log f_{12}/f_{4.6}$ and $\log \lambda_\mathrm{BH}$ is 0.57.

%Note that $\log f_{12}/f_{4.6}$ is used in the RF classifcation, and it may be useful in identifying a small fraction of galaxies with less centrally dense real bulges ($ \log \, \Sigma_1 \lesssim 9.0$). Figure~\ref{fig:wise_sig1} plots $\log \Sigma_1$ versus $\log f_{12}/f_{4.6}$. Real bulges have a wide range of $\log f_{12}/f_{4.6}$. Most of them have low $\log f_{12}/f_{4.6}$ ratios, but a significant fraction of them are still star-forming (see also Luo et al. in prep). 

\begin{figure*}
\includegraphics[width=0.48\textwidth]{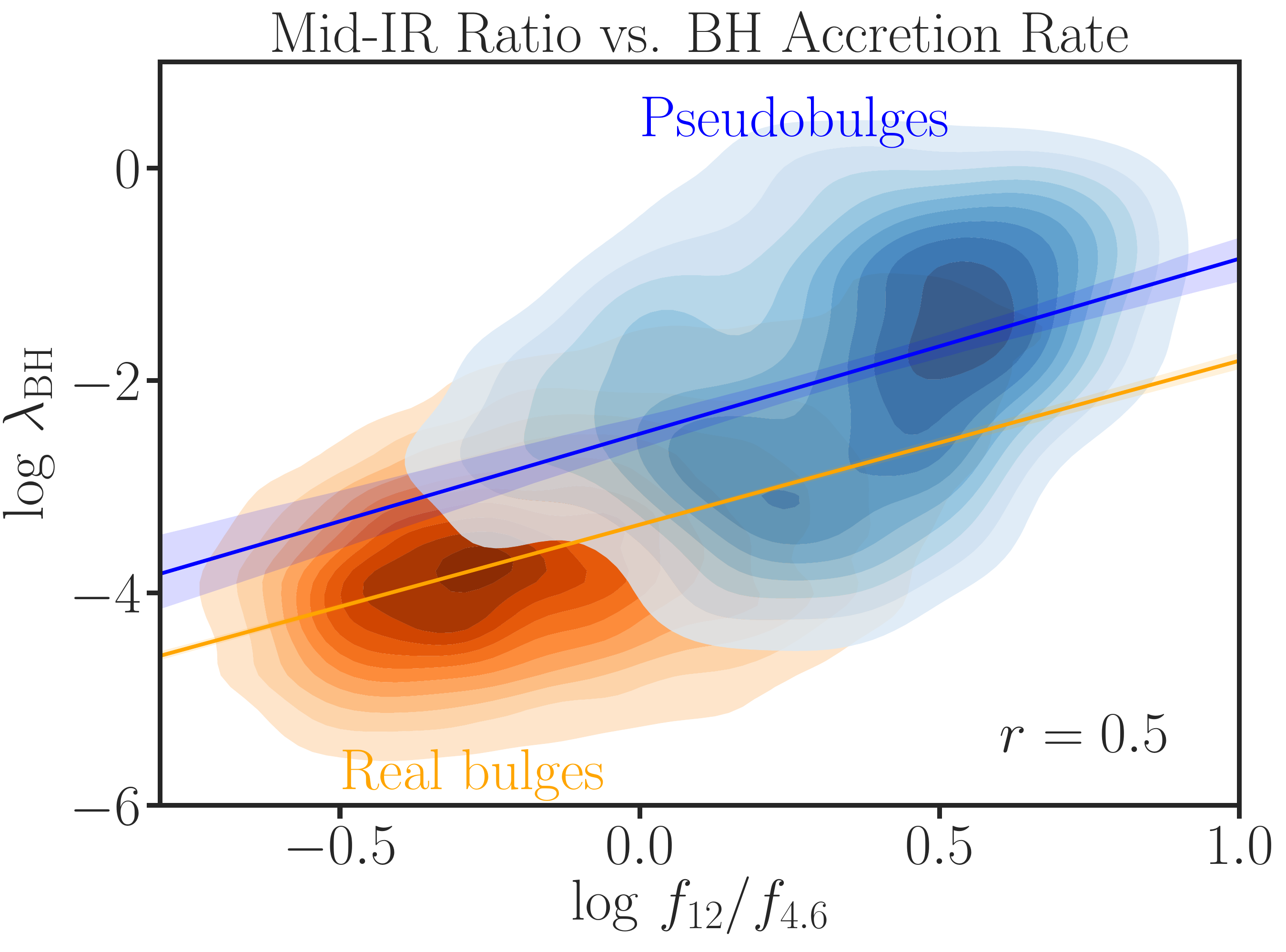}
\includegraphics[width=0.48\textwidth]{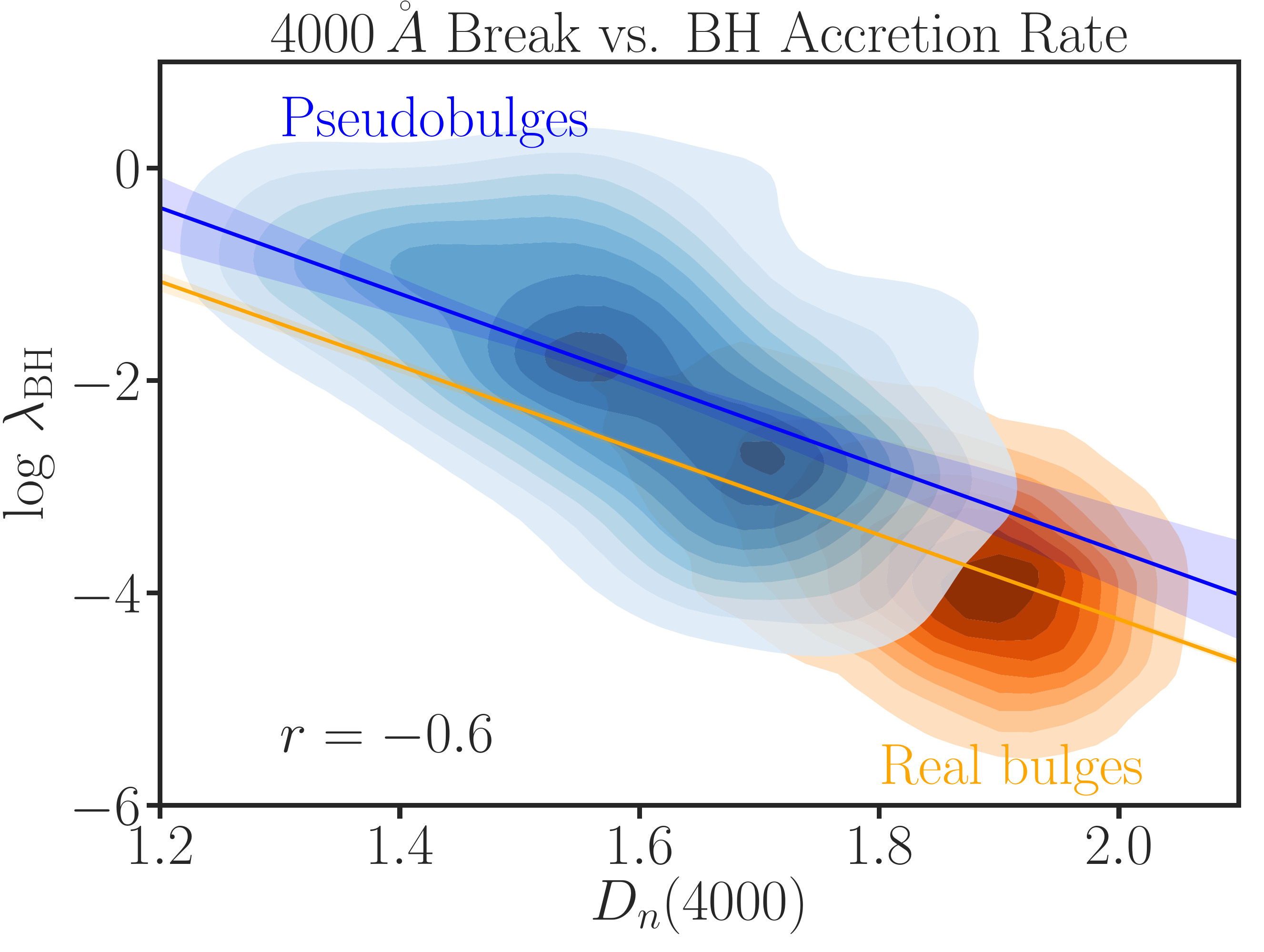}
\includegraphics[width=0.48\textwidth]{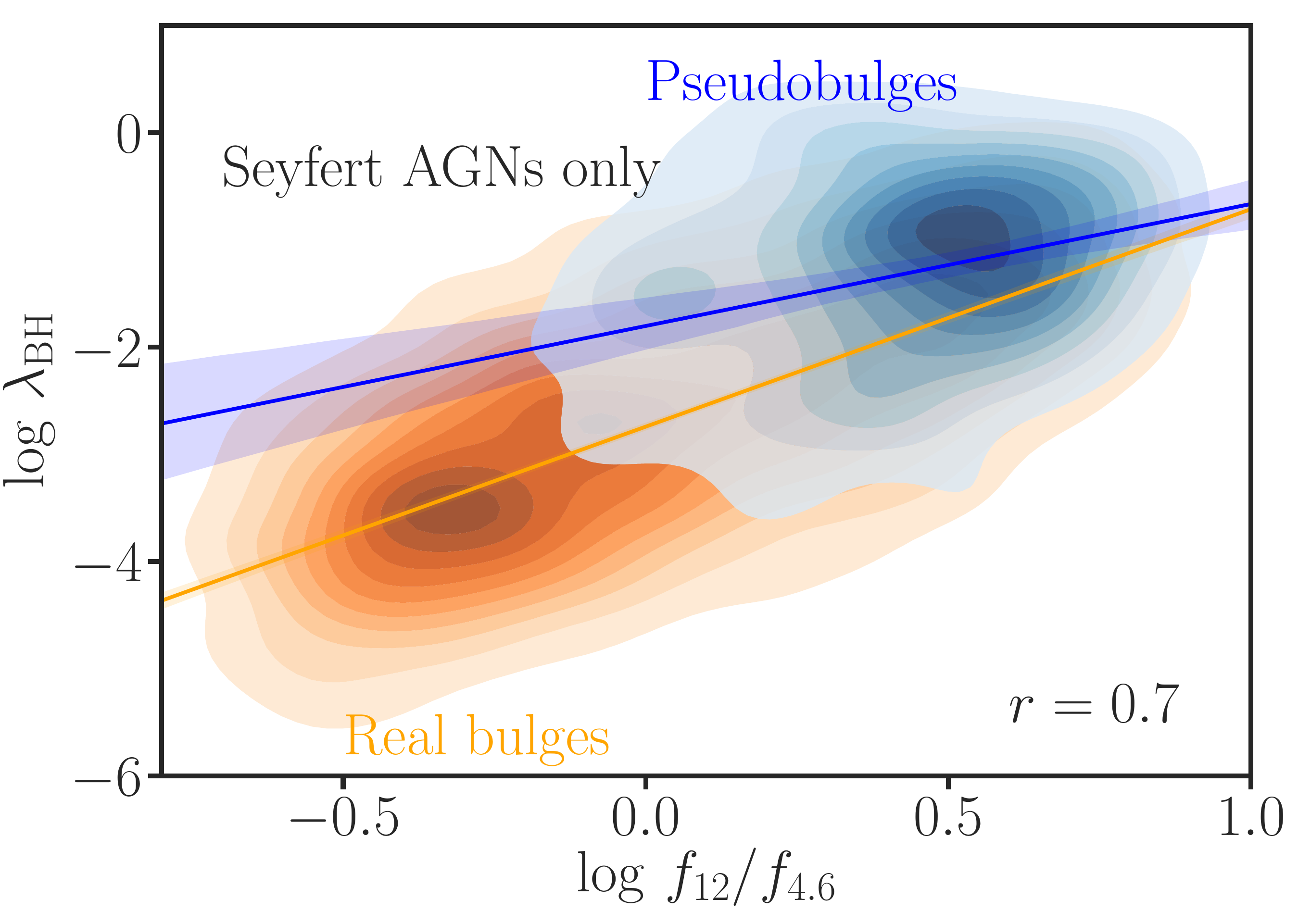}
\hfill
\includegraphics[width=0.48\textwidth]{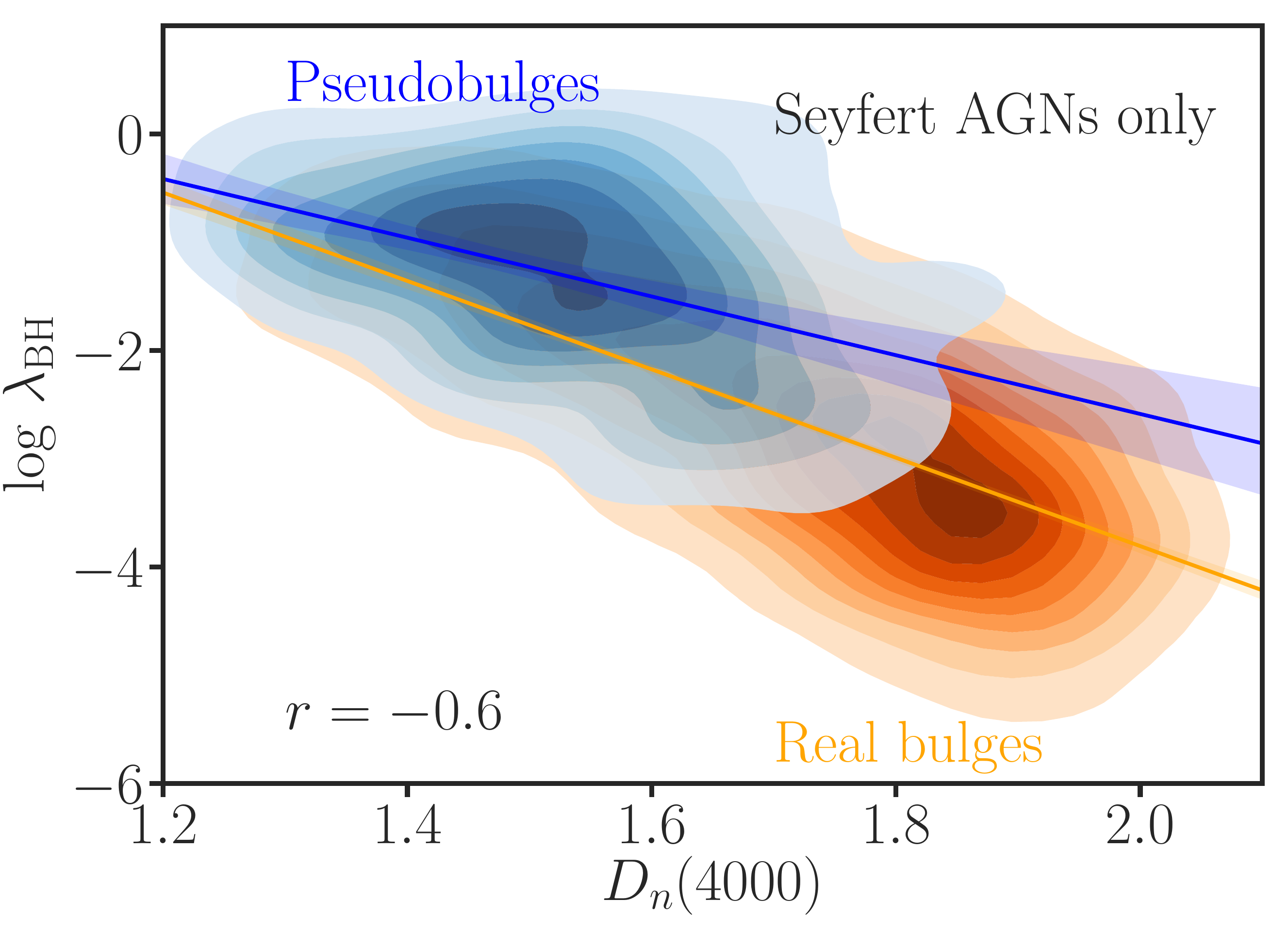}
\caption{The correlation between the WISE flux ratio, $\log f_{12}/f_{4.6}$, or 4000 {\AA} break and the specific black hole accretion rate (Eddington ratio) for pseudobulges and real bulges as defined using the Kormendy relation \citep{Gadotti09}. The $\log f_{12}/f_{4.6}$ ratio is known to correlate with the molecular gas fraction \citep{Yesuf+17}. The top panels include both Seyfert and LINER AGNs while the bottom panels include Seyferts only. On each panel the Pearson correlation coefficient for the combined sample of the two bulge-types is shown. The blue lines and blue shaded regions show the linear regression fits with their 95\% confidence regions for pseudobulges while orange lines show the fits for real bulges. The 95\% confidence regions orange lines are too small to be visible. The figure indicates that both bulge-type and gas fraction likely affect accretion onto black holes.\label{fig:wise_lamBH}}
\end{figure*}

Figure~\ref{fig:wise_lamBH} shows the relationship between $\log f_{12}/f_{4.6}$ or $D_n(4000)$ index and $\log \lambda_\mathrm{BH}$ for AGNs color-coded by the two bulge-types. The top panels show the relation for all AGNs while the bottom panels show the relation for Seyfert AGNs only. The two quantities are correlated/anti-correlated for the AGN samples (the Pearson correlation coefficient is $\sim 0.6$). AGNs hosted by real bulges have a wide range of $\log f_{12}/f_{4.6}$, stellar age and  $\lambda_\mathrm{BH}$. Most of them have low $\log f_{12}/f_{4.6}$ ratios, high $D_n(4000)$, and low specific accretion rates, but some of them are still young and star-forming and have high specific accretion rates. AGNs hosted by pseudobulges have higher star formation rates and higher $\lambda_\mathrm{BH}$. A simple linear regression analysis indicates that the mean relation between $\log f_{12}/f_{4.6}$ or $D_n(4000)$ and $\log \lambda_\mathrm{BH}$ is different for the two bulge-types. Except at high $\log f_{12}/f_{4.6}$ or low $D_n(4000)$ for Seyferts, AGNs in pseudobulges have higher $\log \lambda_\mathrm{BH}$ than those in real bulges. We choose to plot directly observed quantities, $\log f_{12}/f_{4.6}$ and $D_n(4000)$, because the inferred star formation rates are very uncertain. Nevertheless, using the SSFR measurements provided by the SDSS pipeline \citep{Brinchmann+04}, we find that SSFR is also correlated with $\log \lambda_\mathrm{BH}$. The mean relation for all AGNs is described by the linear regression fit $\log \lambda_\mathrm{BH} = 9.0 \pm 0.2 + \mathbbm{1} \{P\} (-3.4 \pm 0.8) + \left [1.06 \pm 0.01 +  \mathbbm{1} \{ P \} (-0.36 \pm 0.07) \right ] \mathrm{SSFR}$, where the indictor function $\mathbbm{1}\{P\}$ is one if an AGN belongs to the pseudobulge class or zero otherwise. The adjusted $R^2$ for the fit is 0.39 (the linear model explains 39\% of the observed variance in $\log \lambda_\mathrm{BH}$). Similarly, the anti-correlation with $D_n(4000)$ is described by the equation $\log \lambda_\mathrm{BH} = 2.9 \pm 0.1 + \mathbbm{1} \{P\} (-0.6 \pm 0.4) + \left [-3.51 \pm 0.05 + \mathbbm{1} \{P\} (0.86 \pm 0.24) \right ] D_n(4000)$ with adjusted $R^2$ of 0.36. The correlation with $\log f_{12}/f_{4.6}$ is described by $\log \lambda_\mathrm{BH} = -3.34 \pm 0.01 + \mathbbm{1} \{P\} (0.84 \pm 0.01) + \left [1.53 \pm 0.03 + \mathbbm{1} \{P\} (0.05 \pm 0.17) \right ] \log f_{12}/f_{4.6}$ with adjusted $R^2$ of 0.3. These trends suggest that the amount of gas in the host galaxies is likely correlated with the specific accretion rate, and the bulge property may modulate how gas is depleted by the star formation and/or how it is accreted on a black hole. The trends do not change qualitatively if we restrict the AGN sample to the Seyferts only.
 %\citep[SSFR,][]{Brinchmann+04} 

In the previous section, we presented the pseudobulge fraction for the galaxies that are AGNs. In Table~\ref{tab:sy_frac}, we present the Seyfert AGN fraction in three SSFR bins for the two bulge-types. Assuming the number of galaxies hosting AGNs is binomially-distributed and approximating the binomial distribution with the normal distribution, since the number of galaxies is large, we estimate the standard errors of the AGN fractions as $\sqrt{ f(1-f)/n}$, where $f$ is the AGN fraction, the number of AGNs divided by the total sample size, $n$, in a given SSFR bin. Although the value of AGN fraction depends on whether LINERs are included or not, the relative AGN fraction at a given stellar mass and SSFR is higher by $\sim 2-3$ times in real bulges than in pseudobulges regardless.

%Note that, at the same SSFR, the pseudobulges have less compact bulges and lower inferred black hole masses than do the overlapping star-forming real bulges. A s
 
To summarize, compared to pseudobulges, real bulges have higher black hole mass, and lower \ion{O}{3} luminosity per black hole mass (specific black hole accretion rate) and lower WISE $\log f_{12}/f_{4.6}$ ratio (specific star formation rate and gas fraction). The $\log f_{12}/f_{4.6}$ ratio and the specific star formation rate significantly correlate with the specific black hole accretion rate.  AGNs hosted by pseudobulges have higher AGNs specific black hole accretions  for the same SSFR than those in real bulges. But the AGN fraction is higher in real bulges.

\section{DISCUSSION}\label{sec:disc}

\subsection{Alternative definition of bulges based on bulge S\'{e}rsic index}

%Namely, the distribution of $n_b$ for 308 nearby galaxies with HST data is bimodal, with a sharp dividing line at $n_b=2$, and with only $\sim 10\%$ overlap between the $n_b$ distributions for classical bulges and pseudobulges. 

Bulges have been defined using the Kormendy relation \citep{Gadotti09} or the bulge S\'{e}risic index, $n_b$ \citep{Fisher+16}. The measurements of the latter using low resolution SDSS images are likely less robust. Training on one definition may not match the classification based on the other perfectly. Therefore, we check that our main results do not change qualitatively if we adopt the bulge-type definition based on $n_b$.

\begin{figure}
\includegraphics[width=0.49\textwidth]{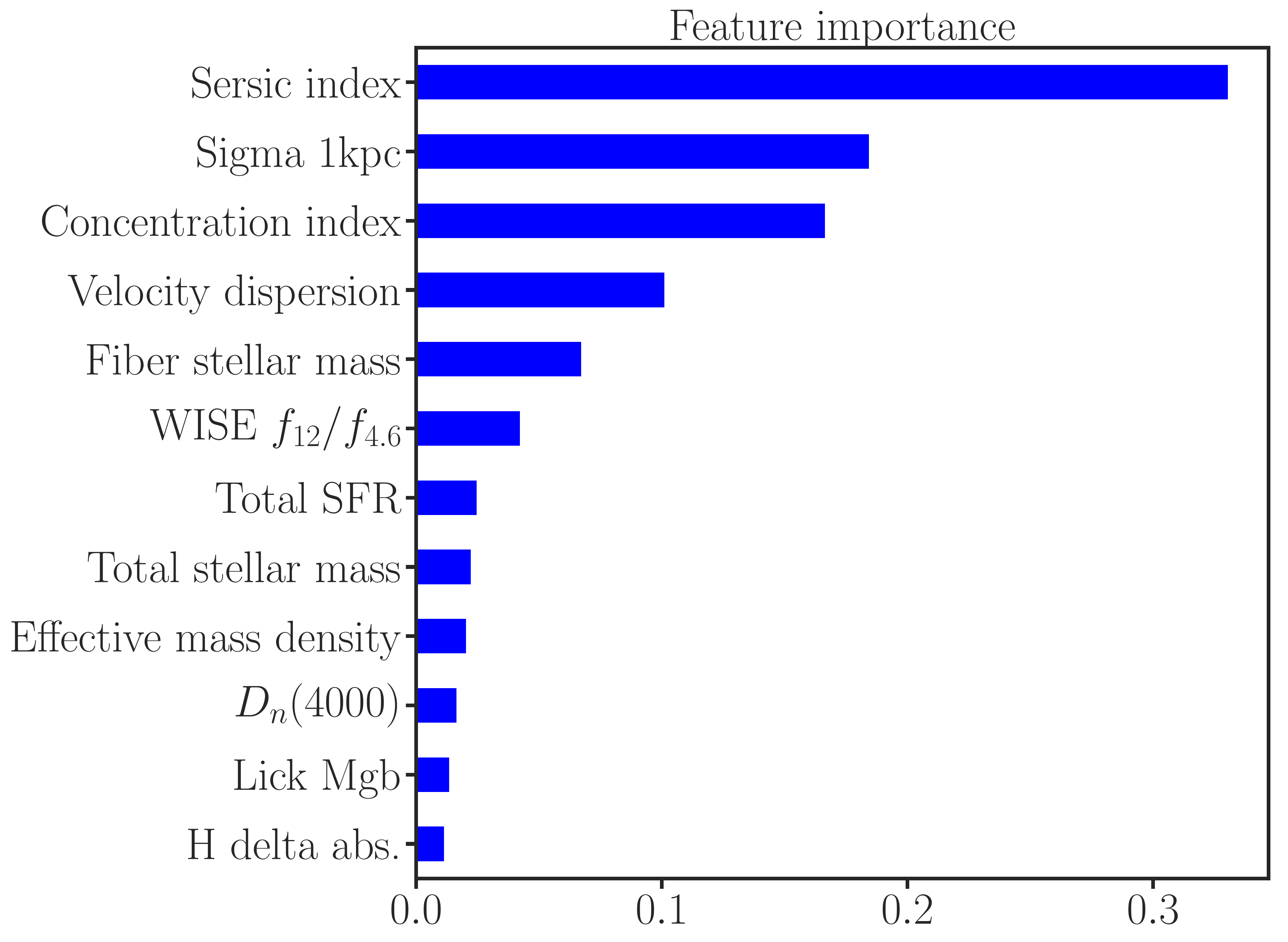}
\caption{The importance of galaxy properties suggested by Random Forest for predicting bulge-types based on the bulge S\'{e}rsic index ($n_b$), as measured by \citet{Gadotti09}. The exact rank of the predictors is not well determined ($\sim1\sigma$) based on the current training data, and the difference from Figure~\ref{fig:imp_nb} is not statistically significant. The combination of the top five variables is useful to accurately predict the bulge-type regardless of how it is defined. \label{fig:imp_nb}}
\end{figure}

\begin{figure*}
\includegraphics[width=0.48\textwidth]{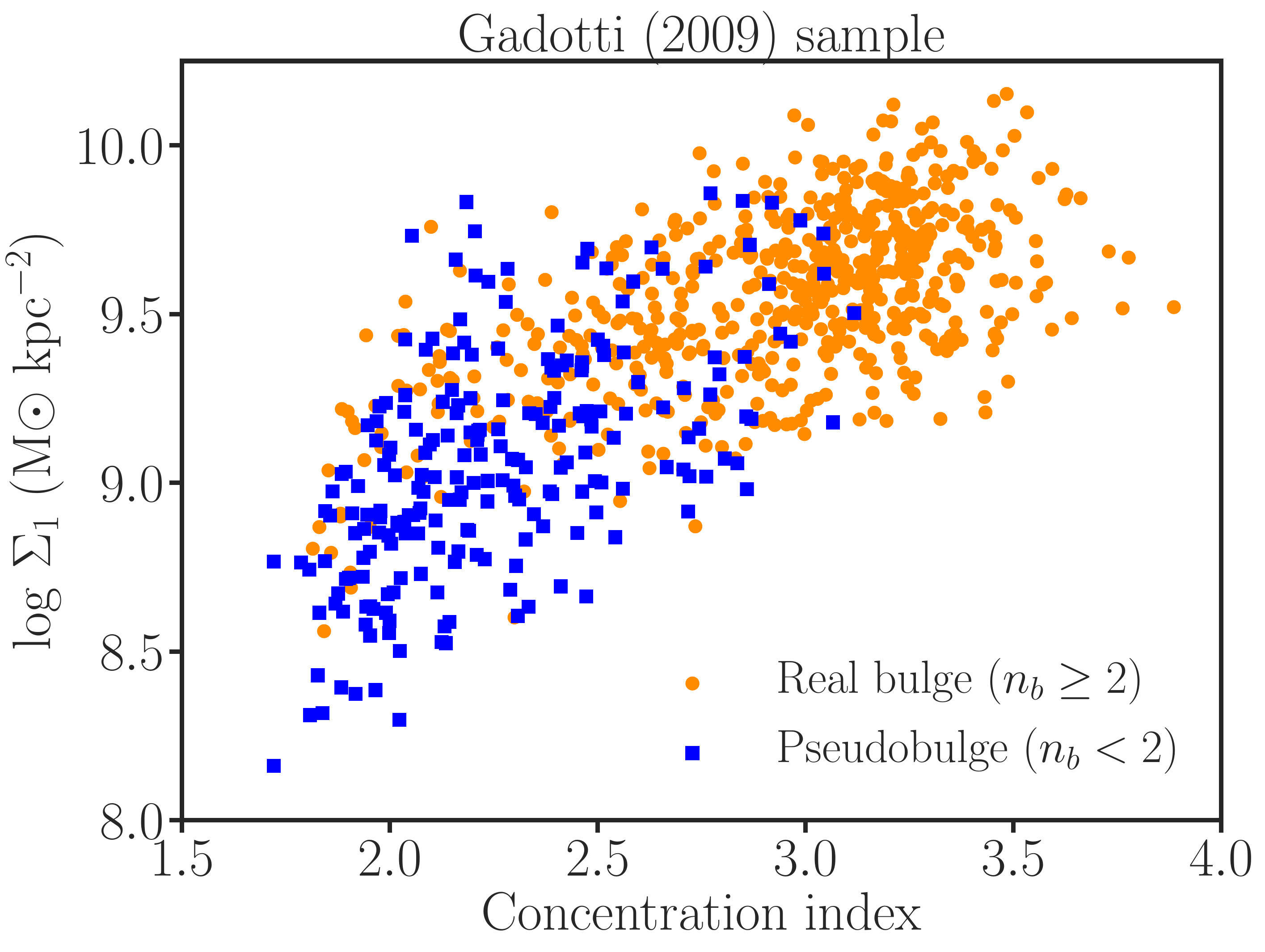}
\includegraphics[width=0.48\textwidth]{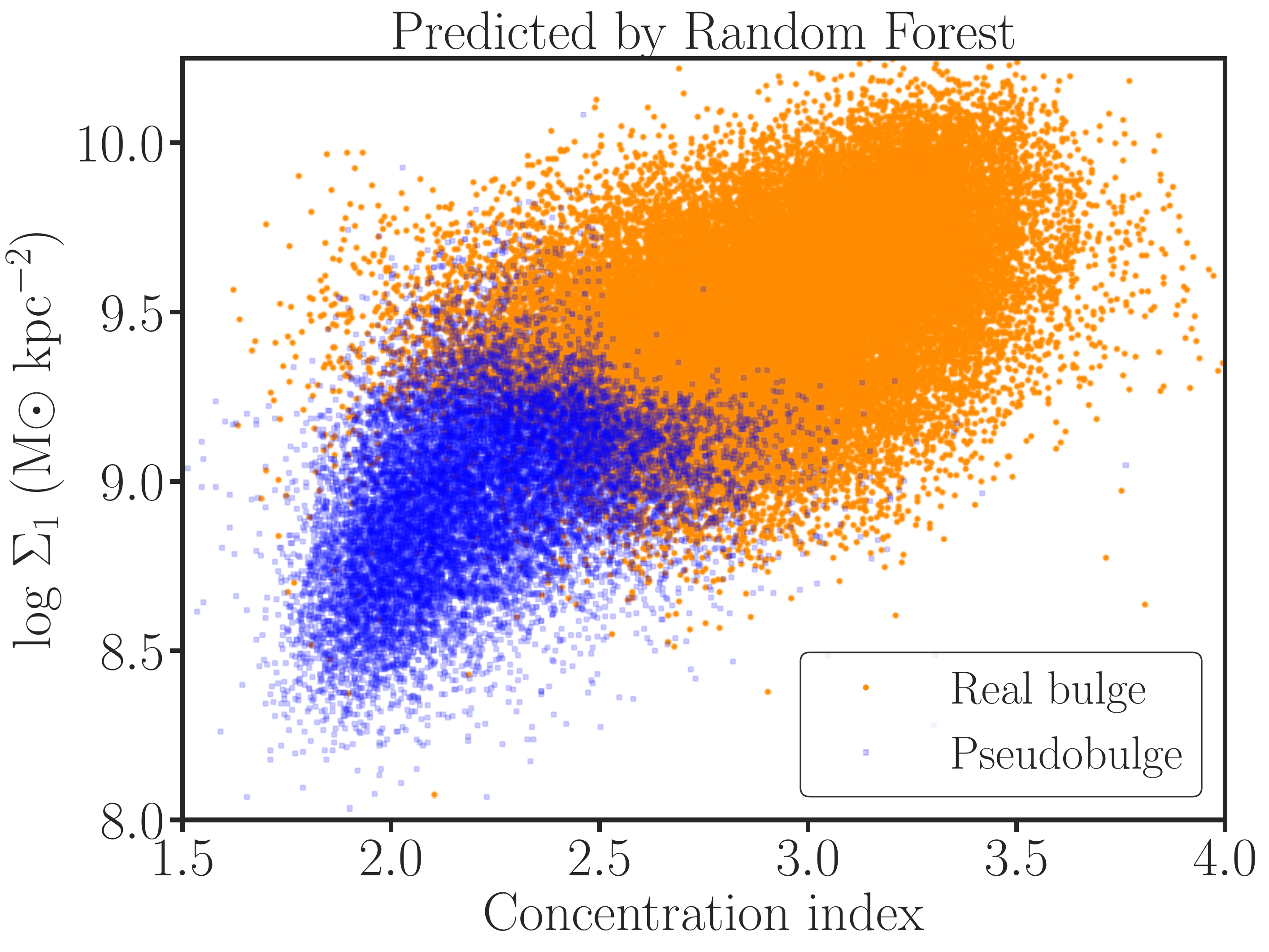}
\includegraphics[width=0.48\textwidth]{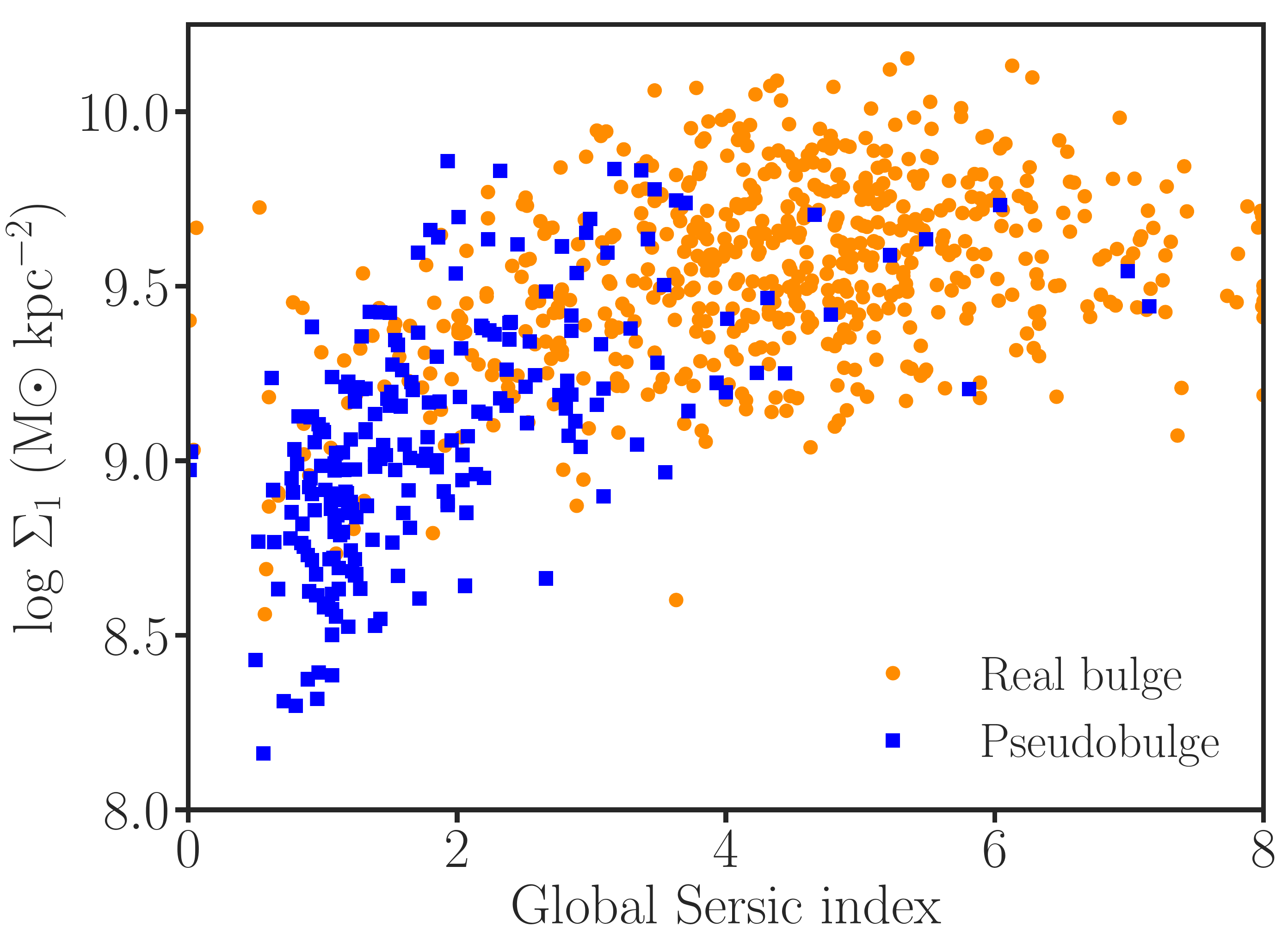}
\includegraphics[width=0.48\textwidth]{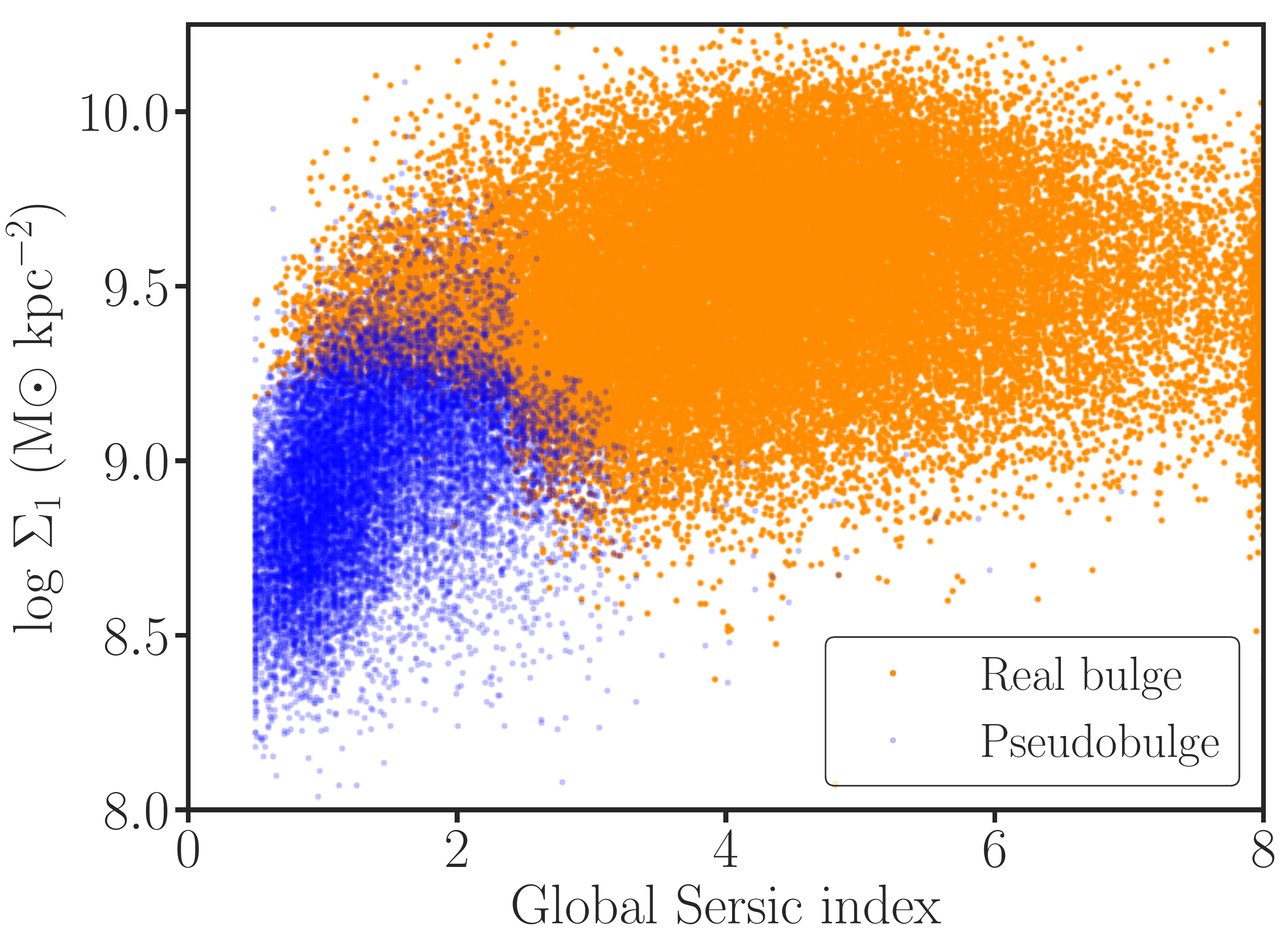}
\includegraphics[width=0.48\textwidth]{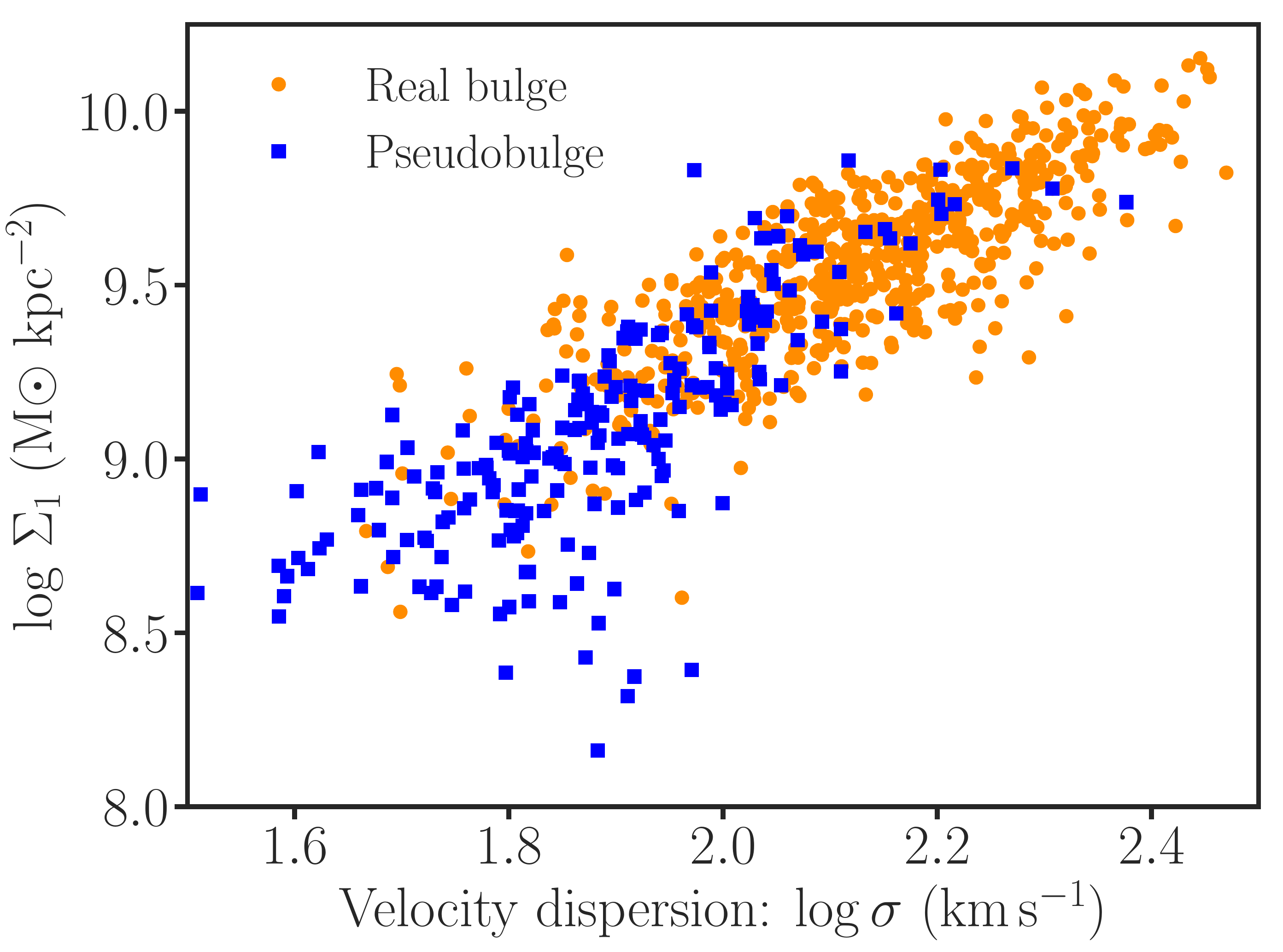}
\hfill
\includegraphics[width=0.48\textwidth]{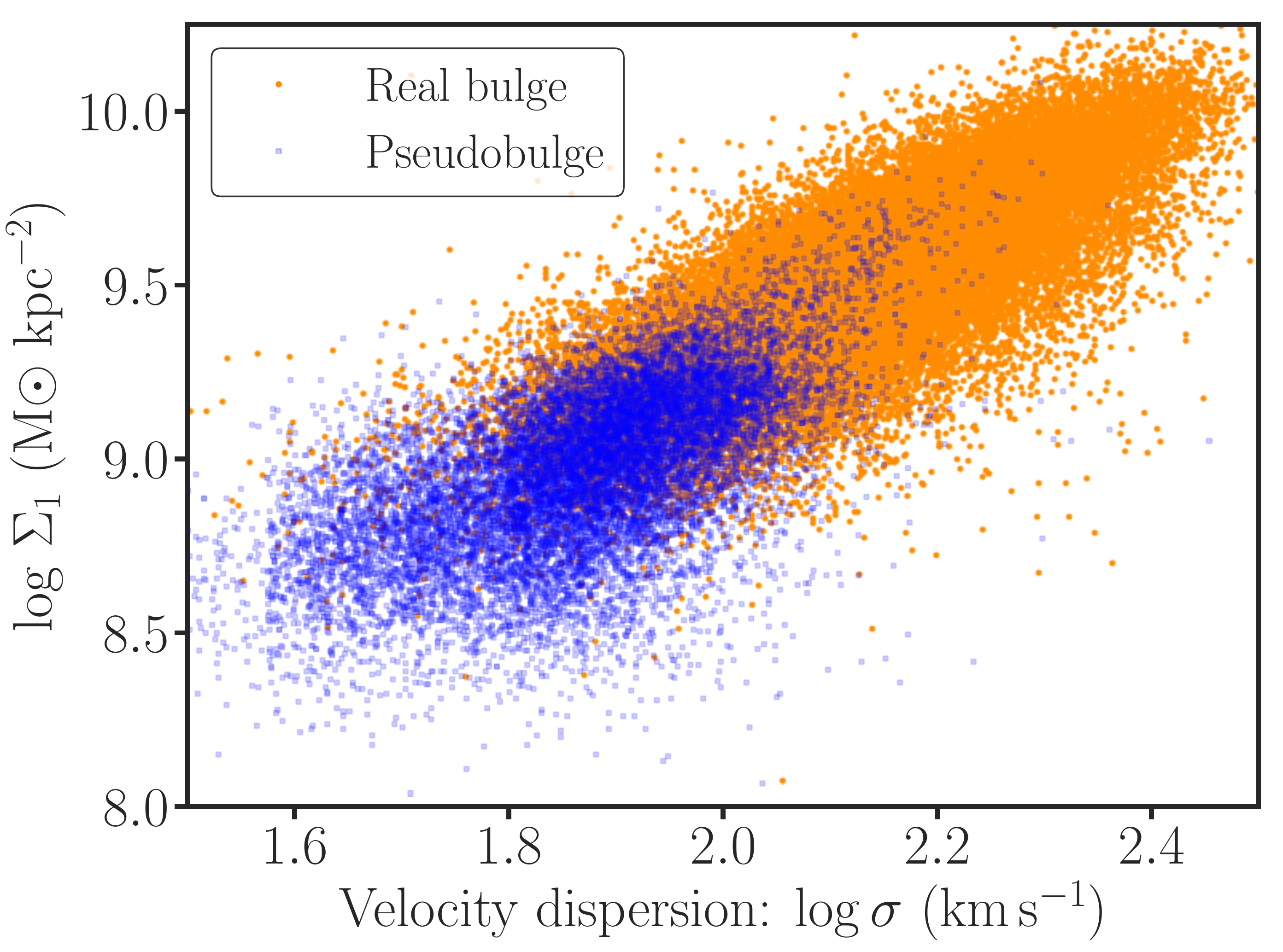}
\caption{Similar to Figure~\ref{fig:train_pred} except here the bulge-types of \citet{Gadotti09}'s sample (left) are based on his measurement of the bulge S\'{e}rsic indices ($n_b < 2$ for pseudobulges). The right panels show the predicted types by Random Forest (RF). The central mass density within 1 kpc, $\Sigma_1$, the global S\'{e}rsic index, and concentration index are among the important parameters suggested by the algorithm. The RF classification maps well the bulge-types from the training sample to the predicted sample. \label{fig:train_pred_nb}}
\end{figure*}

Using \citet{Gadotti09}'s measurements of $n_b$, we divide the training sample into $n_b < 2$ (pseduobulges) and $n_b \ge 2$ (real bulges). The Random Forest algorithm can predict these two categories with $\sim 86 \pm 5\%$ training accuracy, and $\sim 85$\% test accuracy. The RF algorithm correctly classifies 108 galaxies as real bulges and 29 galaxies as pseudobulges and it misclassifies 16 galaxies as real bulges and 9 as pseudobulges, using the same parameters as Section~\ref{sec:RF}. Table~\ref{tab:nb_frac}, Figures~\ref{fig:imp_nb}, and ~\ref{fig:train_pred_nb} repeat the analysis presented in previous sections. The two bulge definitions give similar results. Namely, most pseudobulges have global S\'{e}risic indices $n \lesssim 2.5$, concentration indices $C_r \lesssim 2.5$, velocity dispersions $\sigma \lesssim 100$ \kms\, and central mass densities $\log\,\Sigma_1 \lesssim 9$\,M$_\odot$ kpc$^{-2}$. For the $n_b$ definition, the boundaries for the last three quantities are less clear-cut, and it has relatively more galaxies outside these approximate thresholds.

%\begin{figure*}
%\includegraphics[width=0.48\textwidth]{wise_lamb_BH_all_nb.pdf}
%\includegraphics[width=0.48\textwidth]{wise_lamb_BH_sy_nb.pdf}
%\caption{The correlation between the WISE flux ratio, $\log f_{12}/f_{4.6}$, and the specific black hole accretion rate for pseudobulges and real bulges defined using the bulge S\'{e}rsic index of the training sample.  The $\log f_{12}/f_{4.6}$ ratio is know to correlate with the molecular gas fraction \citep{Yesuf+17}. \label{fig:wise_lamBH_nb}}
%\end{figure*}

%We caution that he $n_b$ measurements of \citet{Gadotti09}'s sample may not very reliable since the centers of majority of these SDSS galaxies are not well resolved. %There are also discrepancies between $n_b$ measurements of \citet{Gadotti09} and \citet{Simard+11} for some objects.

%Furthermore, using $\Sigma_1$ may only marginally improve the classification accuracy. For the classification based Kormendy relation \citet{Gadotti09}, the accuracy on the test set drops from 95.5\% to 93\% if we repeat the classification without using $\Sigma_1$. The accuracy does not change for the classification based on $n_b$. Note that we keep the Random Forest hyper-parameters the same for the analysis with and without $\Sigma_1$. 

\subsection{The $\Sigma_1-M_{BH}$ correlation}

The simple exercise of comparing the radiative accretion energy of supermassive black holes with the binding energy of their host galaxies leads to the conclusion that black holes are energetically viable in clearing-out gas in galaxies \citep[e.g.,][]{Fabian12}. It is thought that AGN feedback affects properties of galaxies (e.g., morphology and star formation rate) and establishes the observed tight correlations of black hole mass with host galaxy stellar properties such as bulge mass and velocity dispersion \citep[e.g.,][]{Kormendy+11,Kormendy+13}. \citet{Kormendy+11} found that black holes correlate differently with different bulge-types. Motivated by this observation, they proposed two different black hole feeding mechanisms: (1) black hole in real bulges grow rapidly when dissipative mergers drive gas to galaxy centers and power bright AGNs. (2) In contrast, small black holes in disk-dominated pseudobulge galaxies grow secularly and power weak AGN activities with little energy to affect their host galaxies.

Based on the correlation between $\Sigma_1$ and velocity dispersion scaled to 1 kpc, $\Sigma_1\propto \sigma_1^{1.99 \pm 0.22}$, \citet{Fang+13} predicted for $z \sim 0$ galaxies a black hole mass scaling relation of the form $M_\mathrm{BH}\propto \Sigma _1^{\alpha/2.0}$, assuming $M_\mathrm{BH}\propto \sigma^{\alpha}$. If this relation is true at all times, $\Sigma_1$ is an easily measurable surrogate for the black hole mass in a host galaxy with a resolved photometry ($< 1$ kpc). As shown in Table~\ref{tab:sig_sig}, using only real bulges from our new classification and different regression methods, we find that $\Sigma_1\propto \sigma_1^{1.44 - 1.72}$, and using $\alpha = 4.38$ \citep{Kormendy+13} gives $M_\mathrm{BH}\propto \Sigma _1^{2.5 - 3.0}$.  All regression methods indicate that the relationship between $\Sigma_1$ and $\sigma_1$ is slightly different when the fit includes pseudobulges. Therefore, caution should be exercised when inferring black hole masses from $\Sigma_1$ when the bulge-type information is not taken into account. Given their estimated error of 0.22, \citet{Fang+13}'s estimate of the slope of the $\log \Sigma_1-\log \sigma_1$ relation is consistent with our estimate when our fit includes pseudobulges. In conclusion, $\Sigma_1$ may be used as an approximate estimator of a black hole mass.

%by simply fitting $\log \Sigma_1$ versus $\log \sigma_1$ and  $\log \sigma_1$ versus $\log  \Sigma_1$ and then taking the geometric mean of the first slope and the reciprocal of the second slope. This method is known as the reduced major axis (RMA) regression. We get $\log \Sigma_1 \propto 1.45 \log \sigma_1$, $\log \sigma_1 \propto 0.44  \log \Sigma_1$ and a slope of $\sqrt{1.45/0.43} = 1.84$. The two fits are shows in Figure~\ref{fig:sig2}. 

%Using both the pseudobulge and real bulge data, the RMA regression gives a slope of 1.87, which is consistent with the estimate of \citet{Fang+13}. The orthogonal regression gives a slope of $2.23 \pm 0.01$ and an intercept of $4.71 \pm 0.01$ when pseudobulges and real bulges are fit together. The intercept increased by 0.2 dex when the two bulge-types are fit together.

\begin{deluxetable*}{lccccc}
\tabletypesize{\footnotesize}
\tablecolumns{6} 
\tablewidth{0pt} 
\tablecaption{The fiber effect on the pseudobulge fraction, as quantified by the change in the fraction at different redshifts. \label{tab:z_pfrac}}
\tablehead{
		\colhead{Redshift} & \colhead{All AGN} & \colhead{Seyfert AGN} & \colhead{AGN+SF Composite} & \colhead{Star-forming (SF)} &  \colhead{All}}
		\startdata
		$z = 0.02-0.03$ & 0.10\ (0.08, 0.14) & 0.15\ (0.11, 0.19) & 0.23\ (0.19, 0.28) & 0.48\ (0.43, 0.53) & 0.21\ (0.18, 0.25) \\ 
	         $z= 0.02-0.04$ & 0.07\ (0.05, 0.10) & 0.12\ (0.09, 0.16) & 0.20\ (0.17,  0.24) & 0.52\ (0.47,  0.56) & 0.20\ (0.18, 0.24) \\
		 $z= 0.04-0.07$ & 0.05\ (0.03, 0.08) & 0.09\ (0.07, 0.12) & 0.17\ (0.15, 0.20) &  0.54\ (0.50, 0.58) & 0.21\ (0.18,  0.24) \\ 
		 $z= 0.06-0.07$ & 0.05\ (0.03, 0.08) & 0.09\ (0.07, 0.12) & 0.16\ (0.14, 0.19) &  0.55\ (0.51, 0.59) & 0.22\ (0.19, 0.25) \\ 
		 $z = 0.02-0.07$ & 0.05\ (0.03, 0.08) & 0.09\ (0.07, 0.12) & 0.18\ (0.15, 0.21) & 0.54\ (0.50, 0.58) & 0.21\ (0.18, 0.24) \\
		\enddata
		\tablecomments{The bulge-type is defined according to \citet{Gadotti09}. There is a noticeable fiber effect on the dilution of AGN signature but it is only $\sim 5\%$.}

\end{deluxetable*}

\begin{deluxetable*}{llcccc}
\tabletypesize{\footnotesize}
\tablecolumns{6} 
\tablewidth{0pt} 
\tablecaption{The Seyfert AGN fractions in the two bulge-types. \label{tab:sy_frac}}
\tablehead{
		\colhead{Mass} & \colhead{Type} & \colhead{Low SSFR} & \colhead{Medium SSFR} & \colhead{High SSFR} &  \colhead{All}}
		\startdata
		    & Pseudobulge & 0.109 (0.048, 0.170) & 0.076\ (0.043, 0.108) & 0.015\ (0.010, 0.019) & 0.022\ (0.019, 0.025)\\
All mass & Real bulge & 0.118\ (0.110, 0.127) & 0.149\ (0.129 0.170) & 0.057\ (0.047, 0.068) & 0.109\ (0.105, 0.113) \\ 
	          & Both & 0.118\ (0.109, 0.127)  & 0.136\ (0.119 0.154) & 0.033\ (0.028, 0.038) & 0.083\ (0.080, 0.086) \\ 
	          \hline
		    & Pseudobulge & 0.111\ (0.027, 0.195) & 0.081\ (0.046, 0.115) & 0.015\ (0.011, 0.019) & 0.021\ (0.018, 0.024) \\
 $\log M= 10.0-10.5 $ & Real bulge & 0.186\ (0.162, 0.209) & 0.173\ (0.143, 0.203) & 0.043\ (0.033, 0.053)  & 0.134\ (0.127, 0.141) \\ 
	          & Both & 0.182\ (0.159 0.205) & 0.148\ (0.124, 0.171) & 0.024\ (0.019 0.028) & 0.079\ (0.075, 0.083)\\ 	          
	          \hline
		    & Pseudobulge & 0.000\ (0.000, ---) &  0.069\ (0.000, 0.161) &0.022\ (0.000, 0.051)  & 0.037\ (0.025, 0.050) \\
 $\log M = 10.5-12 $ & Real bulge & 0.091\ (0.083, 0.099) & 0.136\ (0.106, 0.165) & 0.085\ (0.059, 0.111) & 0.093\ (0.088, 0.098)  \\ 
	          & Both & 0.091\ (0.083, 0.099) & 0.132\ (0.104, 0.161)  & 0.074\ (0.052, 0.096) &  0.090\ (0.085, 0.094)\\      
\enddata
	  \tablecomments{Low specific star formation rate is $\log \mathrm{SSFR} <  -11.5 \, \mathrm{yr}^{-1}$, high is $\log \mathrm{SSFR} > -10.5 \, \mathrm{yr}^{-1}$, and medium is in between the two limits. The Seyfert fraction is defined as the number of galaxies in the Seyfert region of the BPT diagram (Figure~\ref{fig:bpt}) divided by the total number of all galaxies that are classifiable by this diagram. The numbers in the parentheses are the 95\% confidence intervals.}
\end{deluxetable*}

\begin{deluxetable*}{lccccc}
\tabletypesize{\footnotesize}
\tablecolumns{6} 
\tablewidth{0pt} 
\tablecaption{The pseudobulge ($n_b < 2$) fraction for bulge-types defined based on bulge Sersic index. \label{tab:nb_frac}}
\tablehead{
\colhead{Mass} & \colhead{AGN}& \colhead{Seyfert} & \colhead{AGN+SF Composite} & \colhead{SF} &  \colhead{All}}
		\startdata
		$\log M =10 - 10.5 $ & 0.23\ (0.18, 0.29) & 0.25\ (0.20, 0.31) & 0.38\ (0.32, 0.44) & 0.64\ (0.57, 0.70) & 0.41\ (0.35, 0.47) \\ 
	         $\log M = 10.5 -12 $ & 0.16\ (0.12, 0.22) & 0.18\ (0.14, 0.24) & 0.22\ (0.17, 0.28) &  0.32\ (0.27, 0.39) & 0.19\ (0.14, 0.24)  \\ 
		All & 0.18\ (0.13, 0.24) & 0.21\ (0.16, 0.27) & 0.31\ (0.25, 0.37) & 0.57\ (0.51, 0.63) & 0.31\ (0.25, 0.37) \\
		\enddata
	%\tablecomment{
\end{deluxetable*}

\begin{figure}
\includegraphics[width=0.49\textwidth]{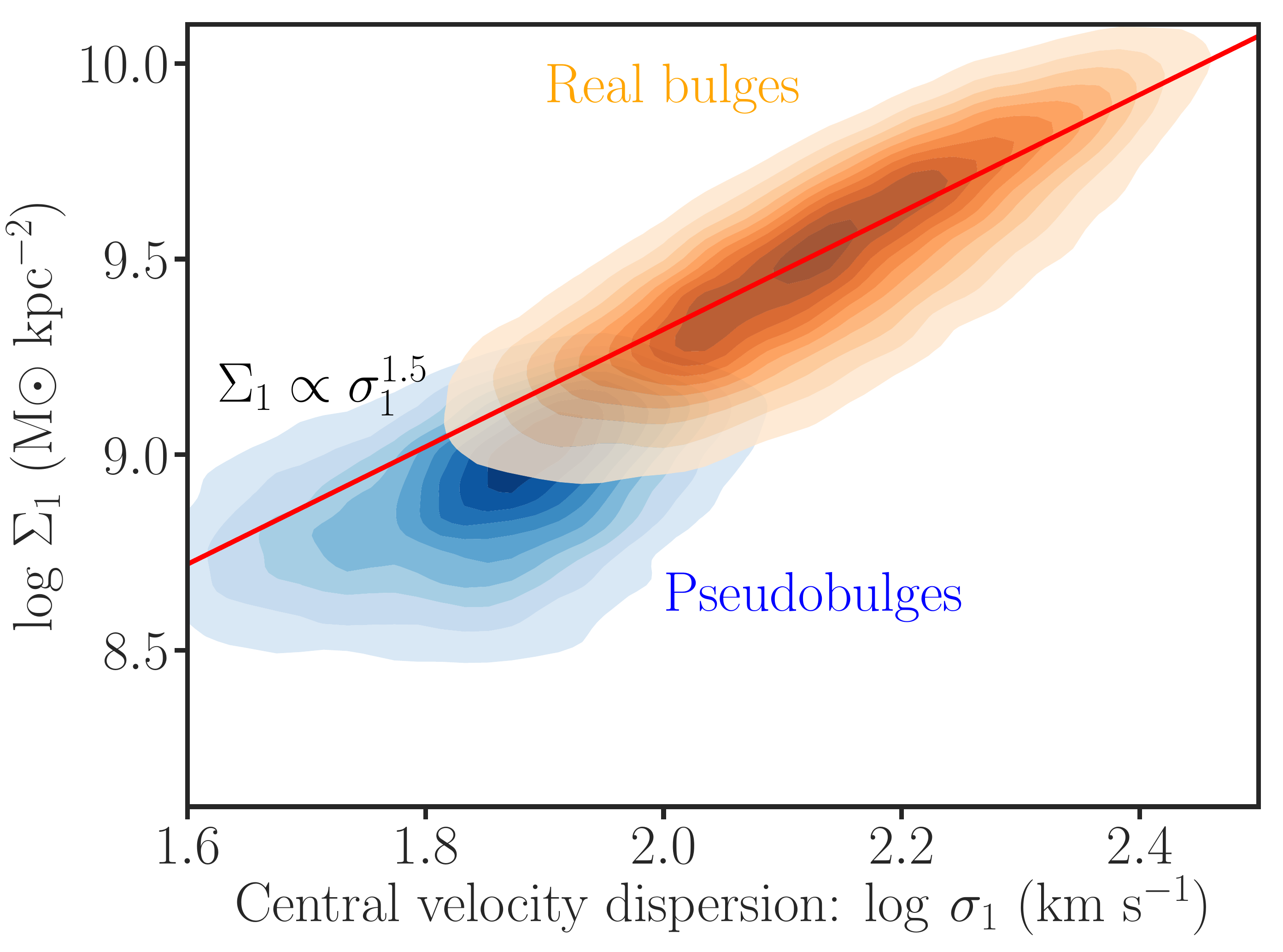}
\caption{The correlation between the central velocity dispersion and the central mass density with 1 kpc. The orange and blue contours show the number density of real bulges and pseudobulges, respectively. The red line is the fit to real bulges only, and it incorporates errors in both quantities. If this relation holds true at all times, $\Sigma_1$ is an easily measurable surrogate for the black hole mass in distant spatially resolved galaxies. \label{fig:sig2}}
\end{figure}

\begin{deluxetable*}{lcc}
\tabletypesize{\footnotesize}
\tablecolumns{2} 
\tablewidth{0pt} 
\tablecaption{Fitting the relation between the central mass density within 1 kpc to the velocity dispersion within 1 kpc,  $\log \Sigma_1 = \alpha+\beta \, \log \sigma_1$, with different regression methods. \label{tab:sig_sig}}
\tablehead{
\colhead{Regression method} & \colhead{Real bulges only}  & \colhead{Both bulge-types}}	
		\startdata
		 Orthogonal Regression  &  $\alpha =  5.84 \pm 0.03$  $\beta = 1.72 \pm 0.01$ & $\alpha =  5.52 \pm 0.01$  $\beta = 1.86 \pm 0.01$ \\ 
		 Simulation Extrapolation (SIMEX) & $\alpha =   6.32 \pm 0.01$  $\beta = 1.49 \pm 0.01$ & $\alpha =   5.77 \pm 0.03$  $\beta = 1.74 \pm 0.01$ \\ 
		 Weighted Median Regression  & $\alpha =  6.35  \pm  0.02$  $\beta =1.49  \pm  0.01$ & $\alpha =  5.98 \pm 0.02$  $\beta = 1.65 \pm 0.01$\\ 
		 Weighted Least Square  & $\alpha =  6.43  \pm  0.01$  $\beta =1.44  \pm  0.01$ & $\alpha =   6.10 \pm 0.01$  $\beta = 1.58 \pm 0.01$ \\ 
	         %OLS Bisector  & $\alpha = 5.61$  $\beta =1.81  \pm 0.01$ \\ 
         	 %Reduced Major Axis  & $\alpha =5.57$  $\beta =1.84  \pm 0.01$\\ 
		 \enddata
	\tablecomments{The top two methods incorporate errors of $\Sigma_1$ and $ \sigma_1$ in the fits, and the last two methods ignore errors of $\sigma_1$. The SDSS  fiber velocity dispersions were  scaled using  the relation $\sigma \propto R^{-0.066}$ \citep{Cappellari+06}.}
\end{deluxetable*}

%\subsection{Theoretical discussion}

%Ho+14 the structure and kinematics of the BLR, as crudely imprinted in the f factor, correlate with Eddington ratio: higher accretion rates result in larger values of f .

\subsection{Comparison with previous work}

\citet{Ho+97} studied the detection rates of emission-line nuclei in the central few hundred parsecs of 420 nearby galaxies, and their dependence on the morphological type and luminosity of the host galaxy. In agreement with our result, they found that  the dominant excitation mechanism of the nuclear emission depends strongly on the Hubble type. Their AGNs reside mainly in early-type (E to Sbc) galaxies, while \ion{H}{2} nuclei prefer late-type (Sbc and later) galaxies. For detailed information of the AGN fractions in different Hubble types, see their Table 2. Their LINER and Seyfert AGN subclasses also have broadly similar host morphologies. 

\citet{Kauffmann+09} reported two modes of black hole growth in SDSS galaxies that are linked to the star formation history and bulge growth. The first mode is associated with young star-forming galaxies. It is characterized by a lognormal distribution of accretion rates that peaks at a few per cent of the Eddington ratio. The second mode has a power-law distribution of accretion rates and is associated with old galaxies with little or no star formation rates. In the lognormal regime, they found that the accretion rate distribution function does not change with galaxy properties such as the black hole mass. They interpreted their result to indicate that black holes regulate their own growth at high gas fractions while their growth is regulated by stellar mass loss at low gas fractions. Our result is qualitatively consistent with \citet{Kauffmann+09} in a sense that most pseudobulge AGNs are young, and have on average high accretion rates while most real bulge AGNs are old and have on average low accretion rates. Detailed comparison with \citet{Kauffmann+09}, however, requires carefully computing the accretion rate distribution functions with possible observational bias corrections, and is beyond the scope of this paper. Moreover, because there is a significant number of young and star-forming real bulges, it will be interesting to study how the accretion rate distributions of young galaxies change relative to their bulge-types. This distinction was not made in \citet{Kauffmann+09}. Figure~\ref{fig:wise_lamBH} suggests that bulge-type correlates with the accretion rate in addition to age and star formation.

\citet{Aird+19} studied the relationship between the star formation rates of galaxies and the incidence of X-ray selected AGNs in distant galaxies from the CANDELS and UltraVISTA surveys out to redshift $z \sim 4$. They also find a linear correlation between the star formation rate and both the AGN fraction and the specific accretion rate. Furthermore, they find that the AGN fraction is significantly elevated for galaxies that are still star-forming but are below the main sequence. Perhaps this enhancement in the AGN fraction is related to the increase in the bulge prominence below the main sequence \citep{Wuyts+11}. %Similarly, \citet{Bernhard+16} found that average starformation rate is  at higher λEdd, as reported in .

Existing cosmological simulations \citep[e.g., Illustris,][]{Sijacki+15} claim to reproduce observed luminosity functions of AGNs, and black hole-host galaxy scaling relations. But they do not have sufficient spatial resolution yet to morphologically distinguish between real bulges and pseudobulges. With a better training sample than currently available, an automated bulge classification of large samples of observed and simulated galaxies will be useful to further constrain theoretical models of bulge formation and galaxy evolution. 

%PB+RB: WLS
 %           Estimate Std. Error t value Pr(>|t|)    
%(Intercept) 6.099885   0.008535   714.7   <2e-16 ***
%Vdisp1      1.581592   0.004155   380.6   <2e-16 ***
%Median Regression
%(Intercept)   5.98458   0.01605  372.91962   0.00000
%Vdisp1        1.64539   0.00793  207.41930   0.00000

\section{SUMMARY AND CONCLUSIONS}\label{sec:conc}

We have shown that one can classify all face-on (axis ratio $b/a > 0.5$) SDSS galaxies above 10$^{10}$ M$_\odot$ and at $0.02 < z < 0.07$, with exception of broadline AGNs, into real bulges or pseudobulges with $\sim 93 \pm 2 $\% accuracy using the Random Forest (RF) algorithm. We use \citet{Gadotti09}'s sample, with his bulge classification labels, as a training and a test sample. Our main conclusions are as follows:

\begin{itemize}

\item When combined, easily measurable structural parameters such as the central mass density with 1\,kpc, the concentration index, the S\'{e}rsic index and velocity dispersion can accurately recover bulge classifications based on image decomposition.

\item AGNs and composite galaxies have lower pseudobulge fractions than do star-forming galaxies. About $75 - 90$\% of AGNs identified by the optical line ratio diagnostic are hosted by galaxies with real bulges. The pseudobulge fraction decreases with the optical AGN line ratio signatures as the ratios change from indicating pure star formation to AGN-dominated in the BPT diagram. The fraction is 54\% (CI: 50--58\%) in the star-forming region, 18\% (CI: 15--21\%) in the AGN+SF composite region and to 5\% (CI: 3--8\%) in the AGN region. After dividing the sample into low and high mass using $\rm{\log M \,(M_\odot) = 10.5}$ as a threshold, the pseudobulge fractions are significantly lower in AGNs than in star-forming galaxies both at high and low masses. 

\item After dividing the sample into three SSFR bins and two bins of stellar mass, we find that the AGN fraction has additional dependance on bulge-types. Even though the precise values of the AGN fraction depend on whether LINERs are included or not as AGNs, the relative AGN fraction in a given stellar mass and SSFR is generally $\sim 2-3$ times higher in real bulges than in pseudobulges.

\item Using dust-corrected \ion{O}{3} luminosity as a proxy for the accretion rate, and the stellar mass and radius as proxies for the black hole mass, we find that, on average, AGNs in real bulges have higher black hole masses but lower \ion{O}{3} luminosities per black hole mass (i.e., lower Eddington ratios) than AGNs in pseudobulges. 

\item The $L_{{\rm O3}}/M_{\rm BH}$ significantly correlates with the WISE $\log f_{12}/f_{4.6}$ ratio, and it anti-correlates with $D_n(4000)$ index. These trends imply correlation with SSFR, which is seen based on the SSFR measurements provided by the SDSS team \citep{Brinchmann+04}. The mean trends between $L_{{\rm O3}}/M_{\rm BH}$ and $\log f_{12}/f_{4.6}$ ratio or $D_n(4000)$ are different for the two bulge-types. Generally, pseudobulges AGNs have higher mean $L_{{\rm O3}}/M_{\rm BH}$ at a given $\log f_{12}/f_{4.6}$ ratio or $D_n(4000)$. 

\item Most pseudobulges do not host AGNs (detectable by optical line ratios) but the ones that host detectable AGNs can have strong AGNs. Therefore, not all pseudobulge galaxies necessarily host weak AGNs.

%Excluding LINERs does not qualitatively change these trends.

\end{itemize}

Furthermore, the following points summarize the discussion presented in Section~\ref{sec:disc} :

\begin{itemize}

%\item We reanalyze the xCOLD GASS sample \citep{Saintonge+18} in the context of our new bulge classification. We find that there is a significant number of real bulges with high gas fractions and SSFRs, similar to those in pseudobulges. These real bulges have higher $\Sigma_1$. There a suggestive evidence that real bulges and pseudobulges may follow different mean relations between the molecular gas fraction and SSFR. Almost all AGNs in the xCOLD GASS survey have real bulge hosts, and they follow a similar mean relation to that of the whole sample. Assuming this mean trend, we infer a correlation between gas fraction and $L_{{\rm O3}}/M_{\rm BH}$.

\item Random Forest recovers the bulge classification based on the bulge S\'{e}rsic index threshold of $n_b = 2$ \citep{Fisher+16} with $\sim 86 \pm 5\%$ training accuracy, and $\sim 85$\% test accuracy. The main conclusions listed above do not change qualitatively if the bulge S\'{e}rsic index is used instead to define the two bulge classes.

\item We revisit the correlation between $\Sigma_1$ and the velocity dispersion scaled to 1 kpc, which \citet{Fang+13} used to predict a black hole mass scaling relation with $\Sigma_1$. We refine their prediction, because we now have the bulge-type information; the black hole masses of pseudobulges do not correlate with the stellar velocity dispersions \citep{Kormendy+13}. Using real bulges only, we find that, depending on a regression model used, $\Sigma_1 \propto \sigma_1^{1.44 -1.72}$, and using $M_{\rm BH} \propto \sigma^{4.38}$ \citep{Kormendy+13} gives $M_\mathrm{BH} \propto \Sigma _1^{2.5-3.0}$.  

\end{itemize}

In future work, the accuracy of the machine bulge-type classification may be improved by including 1) measurements that are derived from image decomposition when they are available, 2) having a large and high resolutions training sample 3) using insights from high resolutions hydrodynamic simulations and 4) combining the RF algorithm with other algorithms.

%Our results unveil the dependence of black hole accretion rate and AGN fraction on the star formation rate (cold gas fraction) and bulge morphology.

\bigskip

We thank the anonymous referee for helpful comments and suggestions. We are grateful to Aldo Rodriguez-Puebla, Kohei Inayoshi, and Luis C. Ho for useful discussion. 

%Funding for SDSS-III has been provided by the Alfred P. Sloan Foundation, the Participating Institutions, the National Science Foundation, and the U.S. Department of Energy Office of Science. The SDSS-III web site is http://www.sdss3.org/.

%SDSS-III is managed by the Astrophysical Research Consortium for the Participating Institutions of the SDSS-III Collaboration including the University of Arizona, the Brazilian Participation Group, Brookhaven National Laboratory, Carnegie Mellon University, University of Florida, the French Participation Group, the German Participation Group, Harvard University, the Instituto de Astrofisica de Canarias, the Michigan State/Notre Dame/JINA Participation Group, Johns Hopkins University, Lawrence Berkeley National Laboratory, Max Planck Institute for Astrophysics, Max Planck Institute for Extraterrestrial Physics, New Mexico State University, New York University, Ohio State University, Pennsylvania State University, University of Portsmouth, Princeton University, the Spanish Participation Group, University of Tokyo, University of Utah, Vanderbilt University, University of Virginia, University of Washington, and Yale University.

\appendix

\section{Separating classical bulges and ellipticals}

\begin{figure*}
\includegraphics[width=0.49\textwidth]{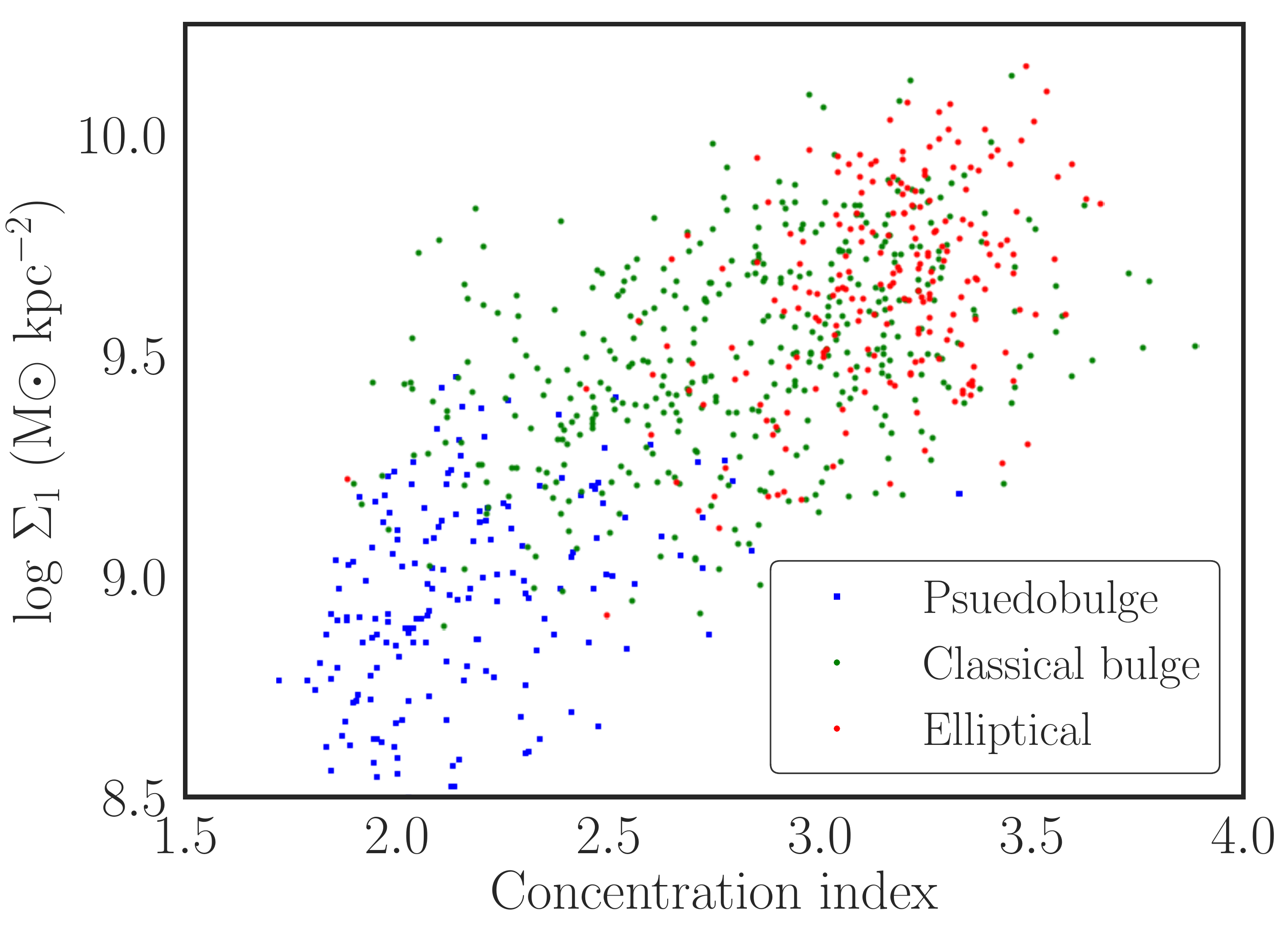}
\includegraphics[width=0.49\textwidth]{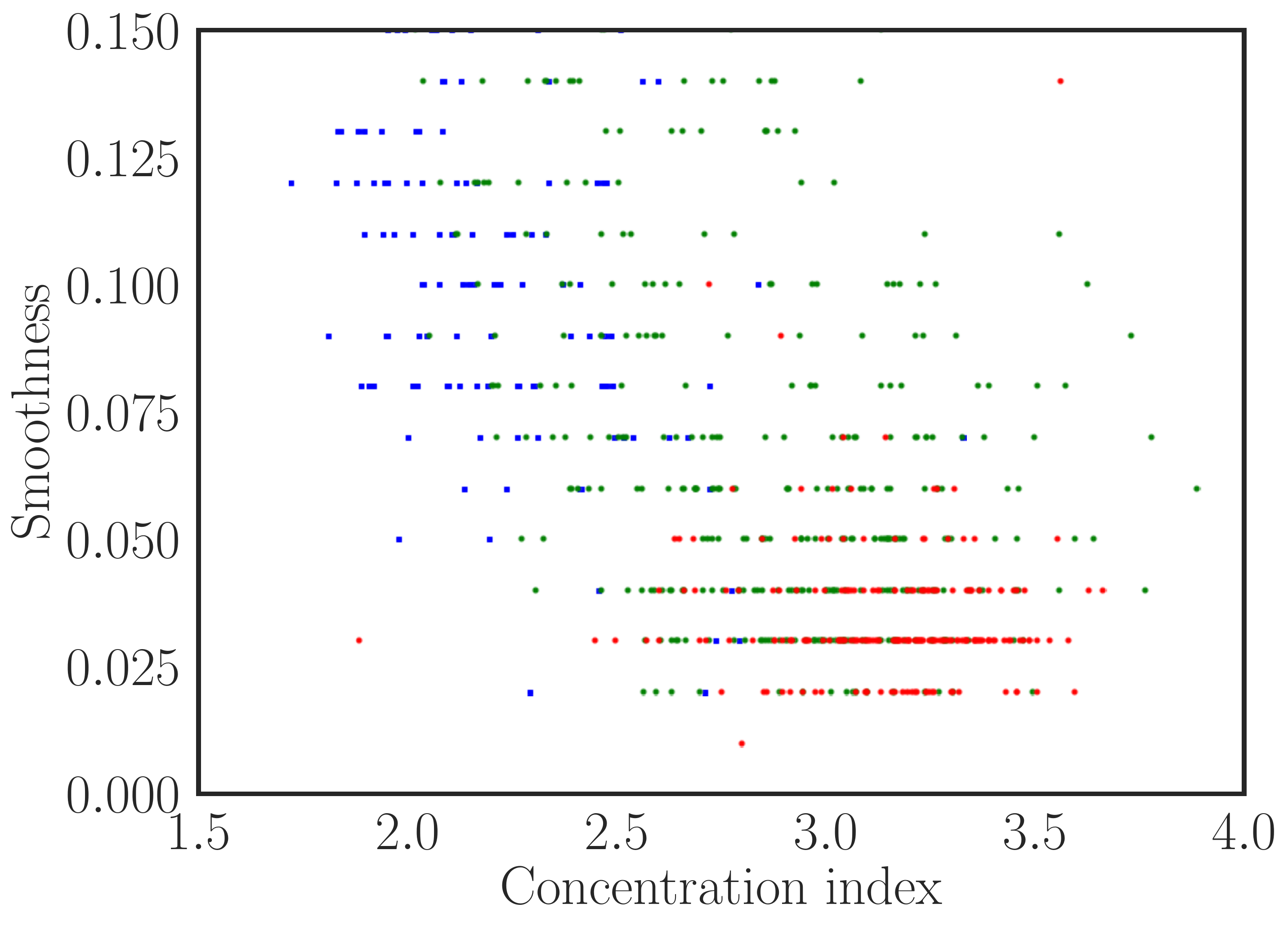}
\includegraphics[width=0.49\textwidth]{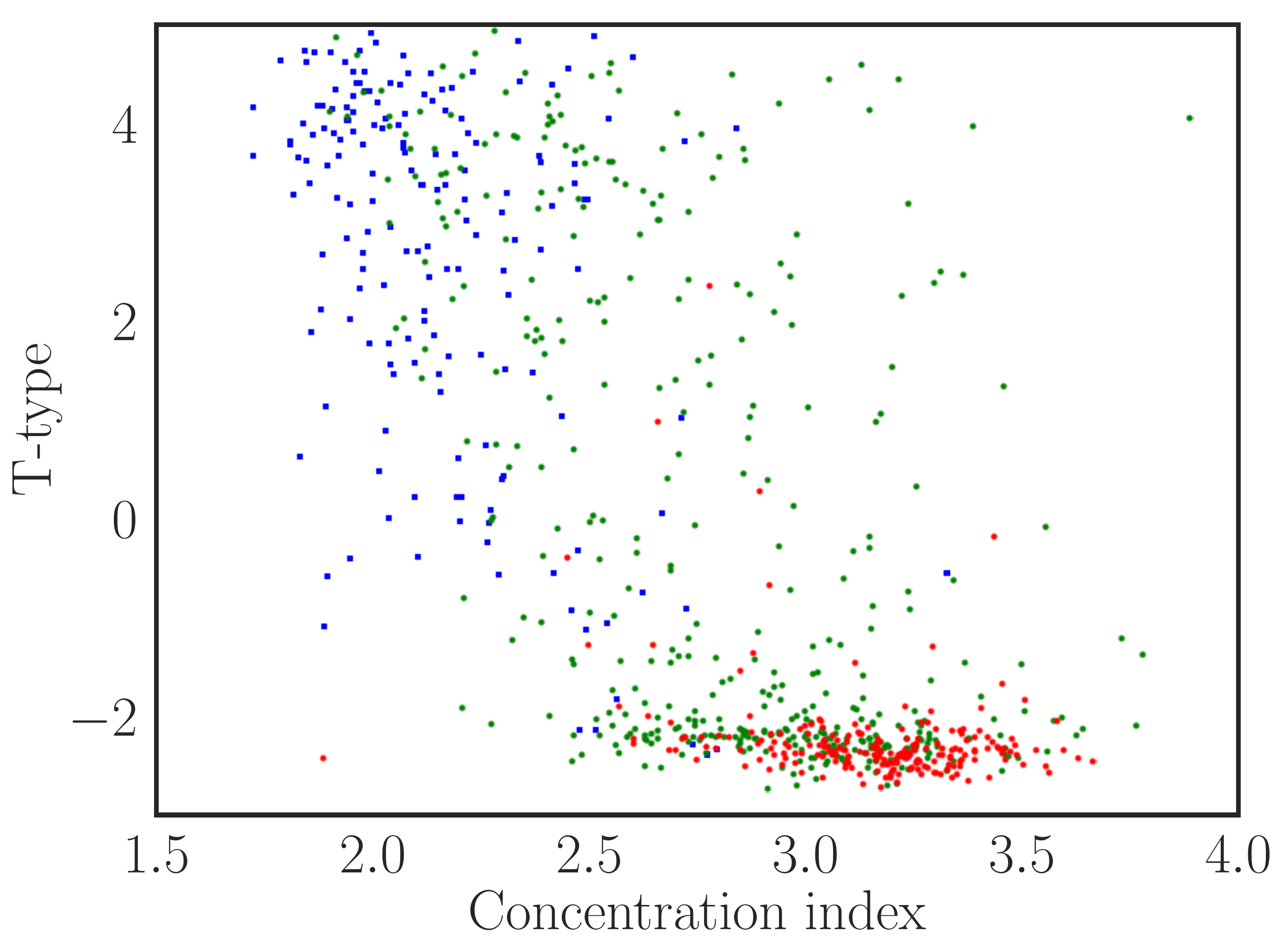}
\hfill
\includegraphics[width=0.49\textwidth]{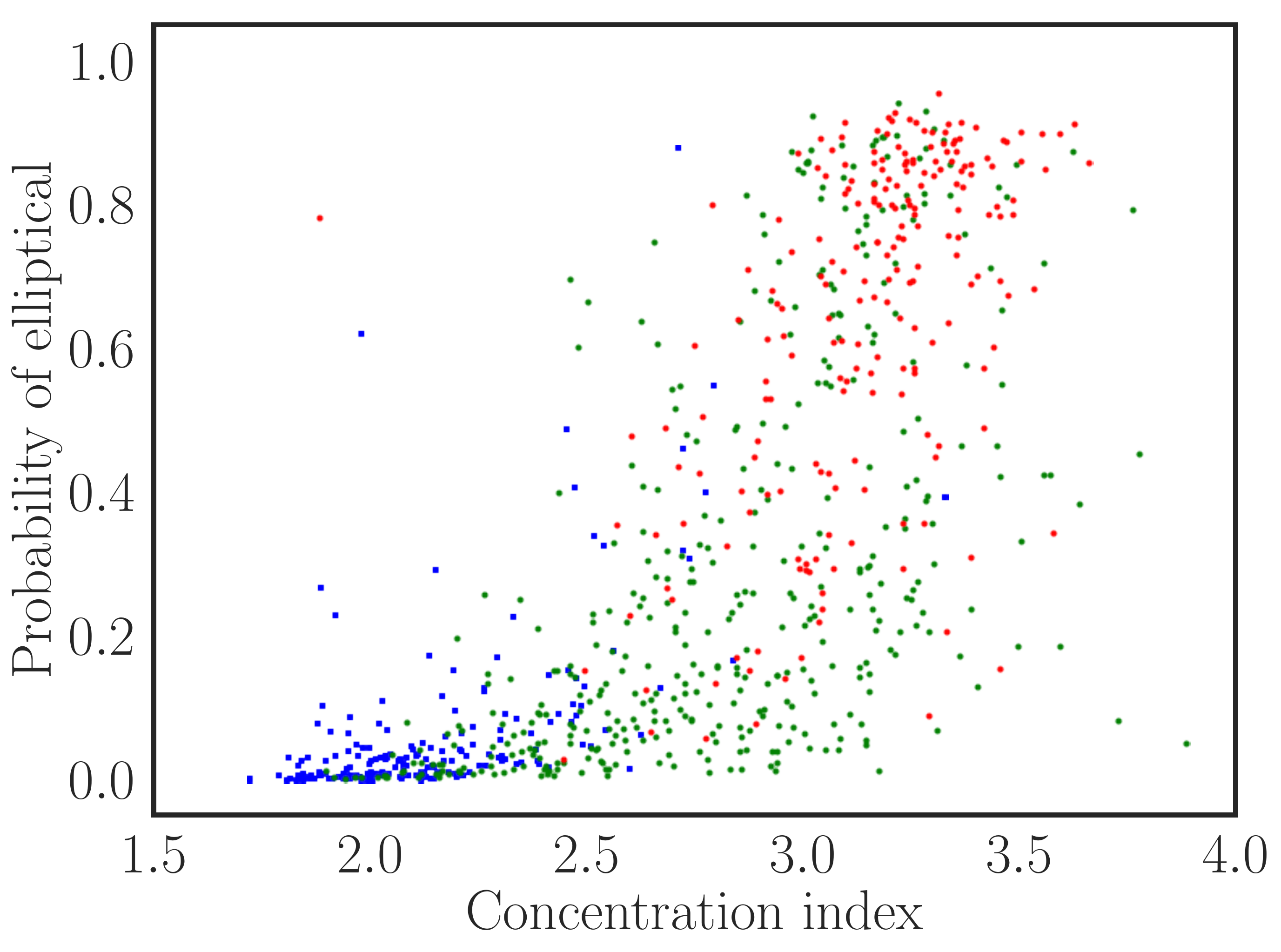}
\caption{Approximate ways of classifying pseudobulges, classical bulges and ellipticals in \citet{Gadotti09}'s sample. Pseudobulges have concentration indices $C_r \lesssim 2.5$  and central densities within 1 kpc $\log\,\Sigma_1 \lesssim 9$ M$_\odot$ kpc$^{-2}$. Classical bulges and ellipticals have significant overlap. Ellipticals classified by \citet{Gadotti09} have on average high concentration indices ($C_r  \lesssim 3$), are smooth ($S2  \lesssim 0.05$), and have T-type $\lesssim -2$. Some classical bulges have $C_r \lesssim 3$ and some of those that have $C_r \gtrsim 3$ have T-types $\gtrsim -2$ and $S2 \gtrsim 0.05$. \label{fig:3class}}
\end{figure*}

The training sample is too small to classify accurately galaxies into three classes: pseudobulges, classical bulges and ellipticals. In the main text, we group the latter two classes together and called them real bulges. The accuracy we achieve for the three class classification using Random Forest is about 75\%. Figure~\ref{fig:3class} shows that combining morphological parameters such as concentration index, $g$-band smoothness/clumpiness parameter \citep[$S2$,][]{Simard+11}, T-type, and the probability of being elliptical \citep{Dominguez+18} can help crudely discriminate between classical bulges and ellipticals. Ellipticals classified by \citet{Gadotti09} have on average high concentration indices ($C_r \gtrsim 3$), are smooth ($S2  \lesssim 0.05$), and have T-types $\lesssim -2$. Some classical bulges have $C_r  \lesssim 3$ and some of those that have $C_r \gtrsim 3$ have T-types $\gtrsim -2$ and $S2 \gtrsim 0.05$. Future work may develop more accurate methods to classify ellipticals and classical bulges.
 
%\bibliography{reference}

%\end{thebibliography}

\end{document}